\documentclass[fleqn]{article}

\usepackage{amsmath,amssymb}
\usepackage{graphicx}
\usepackage[noblocks]{authblk}
\usepackage[top=0.75in, bottom=0.75in, left=0.75in, right=0.75in, dvips]{geometry}
\usepackage{hyperref}
\usepackage[small]{caption}

\def\tref#1#2#3{{#1} (#2) #3}
\newcommand{\ii}{\ensuremath{\mathrm{i}}}
\newcommand{\e}{\ensuremath{\mathrm{e}}}
\newcommand{\ppi}{\ensuremath{\mathrm{\pi}}}
\newcommand{\dif}[1]{\ensuremath{ \mathrm{d}\,#1 }}
\newcommand{\borel}{\ensuremath{\mathcal{B}}}

\newcommand{\qcd}{\ensuremath{\mathrm{QCD}}}

\newcommand{\double}[2]{\ensuremath{(#1,\,#2)}}
\newcommand{\order}[2]{\ensuremath{\mathcal{O}\!\left(#1^{#2} \right)}}
\newcommand{\alpi}{\ensuremath{\frac{\alpha}{\pi}}}

\begin{document}

\title{{\sf Near-Maximal Mixing of Scalar Gluonium and Quark Mesons: A Gaussian Sum-Rule Analysis}}

\author[1]{D.\ Harnett} 
\author[2]{R.T.\ Kleiv}
\author[3]{K.\ Moats}
\author[2]{T.G.\ Steele}
\affil[1]{ 
Department of Physics, University of the Fraser Valley,
Abbotsford, BC, V2S 7M8, Canada}
\affil[2]{Department of Physics and
Engineering Physics, University of Saskatchewan, Saskatoon, SK,
S7N 5E2, Canada}
\affil[3]{Department of Physics, Carleton University, Ottawa, ON, K1S 5B6, Canada}

\maketitle

\begin{abstract}
Gaussian QCD sum-rules are ideally suited to the study of mixed states of gluonium  (glueballs) and quark ($q\bar q$) mesons because of their capability to resolve widely-separated states of comparable strength.  The analysis of the Gaussian QCD sum-rules (GSRs) for all possible two-point correlation functions of gluonic and non-strange ($I=0$) quark scalar ($J^{PC}=0^{++}$) currents 
is discussed. For the non-diagonal sum-rule of gluonic and $q\bar q$  currents we show that perturbative and gluon condensate contributions are chirally suppressed compared to non-perturbative effects of the quark condensate, mixed condensate, and instantons, implying that the mixing of quark mesons and gluonium is of non-perturbative origin.   
The independent predictions of the masses and relative coupling strengths from the non-diagonal and the two diagonal GSRs are remarkably consistent with a scenario of two states
with masses of approximately $ 1\;{\rm GeV}$ and  $1.4\;{\rm GeV}$
that couple to significant mixtures of quark and gluonic currents. The mixing is nearly maximal with the heavier mixed state having a slightly larger coupling to gluonic currents than the lighter state.  
\end{abstract}

\section{Introduction}
The interpretation of the nature of the lightest scalar mesons is one of the most fascinating 
problems in hadronic physics.  The plethora of scalar ($J^{PC}=0^{++}$) states below $2\,{\rm GeV}$ \cite{pdg}
cannot be described by a simple $q\bar q$ nonet, a situation indicative of
exotic states such as gluonium (glueballs) or multi-quark ($q\bar q q\bar q$) states  amongst the known scalar mesons.  In the gluonium scenario, two-body decays to pseudoscalars suggest that the $f_0(1370)$, $f_0(1500)$ and $f_0(1710)$ 
contain strong mixtures of gluonium and $q\bar q$ mesons \cite{close}, with the $f_0(1500)$ favoured as the dominant glueball state \cite{f0_1500_interp}. 
Analyses based on chiral Lagrangians \cite{amir,oller} suggest that the  $f_0(1500)$ and $f_0(1710)$
are mainly gluonium states with a small gluonium component of the $f_0(980)$ \cite{amir}.
Other phenomenological approaches present a scenario of mixing between a $1\,{\rm GeV}$ glueball and the $f_0(980)$, $f_0(1500)$ states of a $q\bar q$ nonet  \cite{minkowski}.  Lattice QCD calculations lead to a scalar gluonium state of approximately $1.6\,{\rm GeV}$ with quenched quarks  
\cite{lattice_quench}. However,  with dynamical quarks the mixing with $q\bar q$ states appears to be very strong,  driving the mass of the lightest flavour-singlet meson
down toward $1\,{\rm GeV}$ with tentative identification of an excited state on the order of $1.5\,{\rm GeV}$ \cite{lattice2}. 
The vast literature on mixing of gluonia in QCD sum-rules is reviewed in detail in \cite{narison_review}.
 The key findings of QCD Laplace sum-rules 
are that admixtures of scalar  gluonium and $q\bar q$ ($I=0$) states exist with masses of approximately $1\,\rm{GeV}$ and $1.6\,{\rm GeV}$ \cite{narison_review,narison,ailin,kisslinger}, a conclusion that is also upheld by studies based on Gaussian QCD sum-rules \cite{harnett_glue,harnett_quark,harnett_proc}.  
In particular, the mixing that results from QCD sum-rules is very large \cite{narison_review,narison}, with comparable couplings of these states to gluonic and $q\bar q$ currents \cite{harnett_glue,harnett_quark,harnett_proc}. 

The results of these different approaches suggest that a consistent scenario of $q\bar q$-gluonium mixing is manifested in the scalar hadronic spectrum as two states  on the order of $1\,{\rm GeV}$ and $1.5\, {\rm GeV}$  that couple to a significant mixture of $q\bar q$ and gluonium currents. 
From both the QCD sum-rule and lattice perspectives, this implies that  the non-diagonal correlation function between $q\bar q$ and gluonic currents must be large enough to describe this behaviour.
   However, perturbative contributions to the non-diagonal correlation function (and hence mixing between gluonium and quark mesons) are chirally suppressed \cite{paver,chanowitz,let_glue,narison_book}.  
Non-diagonal correlation functions have been analyzed in detail for pseudoscalar gluonium leading to a small mixing angle 
even in the presence of chiral-violating condensates \cite{paver}; given the similarities in field-theoretical structure between the scalar and pseudoscalar channels, a similarly small mixing angle in the scalar channel seems unavoidable \cite{paver,narison_book}.  However, the mixed condensate effects, which are zero at leading order in the pseudoscalar channel \cite{paver}, are shown below to be non-zero for the scalar channel providing a scale for large mixing.  In addition, we consider instanton effects in our analysis; such effects have been argued to be essential for studies of gluonium \cite{forkel_instanton}. 
As will be seen below (and as argued in \cite{hongying} for glueball decays), the full inclusion of chiral-violating effects of QCD condensates and instantons provide the dominant contributions to the non-diagonal correlator in the scalar channel.  These chiral-violating effects in the non-diagonal correlator are  
 essential for a self-consistent scenario 
of two states coupling to a strong (near maximal) mixture of gluonium and $q\bar q$ currents; this scenario emerges 
from all possible correlation functions of gluonic and $q\bar q$ quark currents ({\it i.e.,} diagonal gluonic, diagonal $q\bar q$, non-diagonal gluonic-$q\bar q$).

The formulation and analysis of Gaussian sum-rules is reviewed in Section~\ref{gsr_sec}.  In Section~\ref{corr_sec}  
the leading-order perturbative, QCD condensate, and instanton contributions to the  non-diagonal correlation function of $q\bar q$ and gluonic currents are calculated along with the associated  Gaussian QCD sum-rules. The analysis of the Gaussian sum-rules and the pattern of state coupling mixing is then presented in Section~\ref{analysis_sec}.

\section{Review of Gaussian Sum-Rules}\label{gsr_sec}
Gaussian sum-rules associated with QCD two-point correlation functions have been shown to be sensitive to the  hadronic spectral functions over a broad energy range, and analysis techniques have been developed to exploit this dependence to determine how resonance strength is
distributed in the spectral function~\cite{harnett_glue,harnett_quark,harnett_proc}.  Thus Gaussian sum-rules are well-suited to situations such as $q\bar q$-gluonium mixing where multiple hadronic states could contribute to a correlation function.

The ($k=0$) Gaussian sum-rule (GSR), introduced in~\cite{gauss}, is given by
\begin{equation}
G_0\left(\hat s,\tau\right)=\frac{1}{\sqrt{4\ppi\tau}} \int\limits_{t_0}^\infty
\exp\left[\frac{-\left(t-\hat{s}\right)^2}{4\tau}\right]\,\frac{1}{\ppi}\rho(t)\;\dif{t}
~,~\tau>0
\label{basic_gauss}
\end{equation}
and relates a QCD calculation $G_0\left(\hat s,\tau\right)$ to a weighted integral of its associated hadronic spectral function
$\rho(t)$ from its threshold $t_0$.  
The Gaussian kernel peaked at $t=\hat s$ smears the spectral function through an (approximate) interval 
$\hat s-2\sqrt{\tau}\le t\le \hat s+2\sqrt{\tau}$.
This smearing provides a clear conceptual implementation
of quark-hadron duality.  The width of this duality interval is constrained from below by QCD because renormalization-group improvement
of $G_0\double{\hat s}{\tau}$ 
sets the renormalization scale $\nu$ through
$\nu^2=\sqrt{\tau}$~\cite{harnett_quark,gauss}; therefore it is not possible to achieve the formal $\tau\to 0$ limit
where complete knowledge of the spectral function could be obtained through
\begin{equation}
\lim_{\tau\to 0}G_0\left(\hat s,\tau\right)=\frac{1}{\ppi}\rho\left(\hat s\right)~,~ \hat s>t_0 ~.
\label{gauss_limit}
\end{equation}

In contrast, the variable $\hat{s}$ in~\eqref{basic_gauss} is unconstrained by QCD and 
can be varied to probe excited and ground states with similar sensitivity.
Any features of the spectral function strong enough to be isolated from the
continuum will be revealed through the GSR.
This behaviour should be compared to that of the Laplace sum-rules
 \begin{equation}
R\left(\Delta^2\right)=\int\limits_{t_0}^\infty\exp{\left(-\frac{t}{\Delta^2}\right)}\frac{1}{\pi}\rho(t)\, \dif{t}
\label{basic_laplace}
\end{equation}
which exponentially suppress excited states relative to the ground 
state.\footnote{The configuration-space correlation function 
in~\eqref{basic_corr_fn} as used in lattice QCD involves exponential suppression  
of excited states similar to that occurring for Laplace sum-rules.}

Sum-rules analyses start from the calculation of an appropriate QCD correlation function
of renormalized composite operators.
We focus on two-point functions of scalar operators $J_1$ and $J_2$
\begin{equation}
  \Pi\left(Q^2\right)=\ii\int\mathrm{d}^4x\;
  \e^{\ii q\cdot x}\left\langle 0 \vert T\left[ J_1(x) J_2(0)\right] \vert 0 \right\rangle
  ~,~ Q^2\equiv -q^2
\label{basic_corr_fn}
\end{equation}
where $J_1$ and $J_2$ can be the same (diagonal) or different (non-diagonal).
The correlator~\eqref{basic_corr_fn} is related to a
hadronic spectral function $\rho(t)$ through a dispersion
relation with a number of subtraction constants. 
For example, in the diagonal scalar gluonic case,
we have
\begin{equation}
  \Pi\left(Q^2\right)-\Pi(0)-Q^2\Pi'(0)-\frac{1}{2}Q^4\Pi''(0)
  =-\frac{Q^6}{\pi}\int\limits_{t_0}^\infty
  \frac{\rho(t)}{t^3\left(t+Q^2\right)}\;\dif{t}
~.
\label{disp_rel}
\end{equation}
Unknown subtraction constants and field-theoretical divergences can be eliminated
by constructing the GSRs\footnote{This definition is a natural generalization of that given in~\cite{gauss}.
To recover the original GSR, we simply let $k=0$ in~(\ref{srdef}).}
\begin{equation}\label{srdef}
   G_k(\hat{s},\tau)\equiv \sqrt{\frac{\tau}{\ppi}}\borel
   \left\{ \frac{(\hat{s}+\ii\Delta)^k \Pi(-\hat{s}-\ii\Delta)
         - (\hat{s}-\ii\Delta)^k \Pi(-\hat{s}+\ii\Delta) }{\ii\Delta}
   \right\}
\end{equation}
where $k\in\{ -1,0,1,\ldots\}$ and where the Borel transform $\borel$ is defined by
\begin{equation}\label{borel}
  \borel\equiv \lim_{\stackrel{N,\Delta^2\rightarrow\infty}{\Delta^2/N\equiv 4\tau}}
  \frac{(-\Delta^2)^N}{\Gamma(N)}\left( \frac{\mathrm{d}}{\mathrm{d}\Delta^2}\right)^N ~.
\end{equation}
Applying~\eqref{srdef} to~\eqref{disp_rel} yields the
following one-parameter family of GSRs (see~\cite{harnett_glue} for further details):
\begin{equation}
 G_{k}(\hat{s},\tau)+ \delta_{k,\,-1}\frac{1}{\sqrt{4\pi\tau}}
 \exp\left( \frac{-\hat{s}^2}{4\tau}\right)\Pi(0)
 = \frac{1}{\sqrt{4\pi\tau}}\int_{t_0}^{\infty} t^k
   \exp\left[ \frac{-(\hat{s}-t)^2}{4\tau}\right]\frac{1}{\pi}\rho(t)\;\dif{t}~.
\label{gauss_family}
\end{equation}
Note that the $k=-1$ sum-rule can only be used in situations where there
exists an appropriate low-energy theorem from which we can determine the
subtraction constant $\Pi(0)$.
Such is the case, for instance, with the diagonal scalar gluonic
two-point function~\cite{let_glue}.

On the right-hand side of~(\ref{gauss_family}), 
we impose a fairly general resonance(s) plus continuum model
\begin{equation}
  \rho(t)=\rho^{{\rm had}}(t)+\theta\left(t-s_0\right){\rm Im}\Pi^{\qcd}(t)
\label{respcont}
\end{equation}
where $s_0$ represents the onset of the QCD continuum.
The resulting continuum contribution 
\begin{equation}\label{continuum}
   G_k^{{\rm cont}} (\hat{s},\tau,s_0) = \frac{1}{\sqrt{4\pi\tau}}   \int_{s_0}^{\infty} t^k
   \exp \left[ \frac{-(\hat{s}-t)^2}{4\tau} \right]  \frac{1}{\pi} {\rm Im} \Pi^{\qcd}(t)\; \dif{t}
\end{equation}
is then moved to the left-hand side of~(\ref{gauss_family}).
The total QCD contribution
\begin{equation}
  G_k^{\qcd}\left(\hat{s},\tau,s_0\right) \equiv G_k\left(\hat{s},\tau\right) -  G_k^{{\rm cont}} \left(\hat{s},\tau,s_0\right)
\label{blah}
\end{equation}
then satisfies
\begin{equation}
 G_{k}^\qcd\left(\hat{s},\tau,s_0\right)+ \delta_{k,\,-1}\frac{1}{\sqrt{4\pi\tau}}
 \exp\left( \frac{-\hat{s}^2}{4\tau}\right)\Pi(0)
  =\frac{1}{\sqrt{4\pi\tau}}\int_{t_0}^{\infty} t^k
   \exp\left[ \frac{-(\hat{s}-t)^2}{4\tau}\right] \frac{1}{\pi} \rho^{\rm had}(t)\; \dif{t}
   ~.
\label{final_gauss}
\end{equation}

Integrating both sides of~\eqref{final_gauss} with respect
to $\hat{s}$ gives
\begin{equation}
  \int\limits_{-\infty}^\infty G_k^{\qcd}(\hat{s},\tau,s_0)\;\dif{\hat{s}}+\delta_{k,\,-1}\Pi(0)
  =\int\limits_{t_0}^{\infty} t^k\frac{1}{\pi}\rho^{{\rm had}}(t)\;\dif{t}~ ,
  \label{tom_norm_2}
\end{equation}
in which the right-hand side is recognized as the $k^{\text{th}}$ member of the finite-energy sum-rule (FESR) family.
Thus, the information contained in the GSRs which is
independent of the FESRs can be isolated by considering the
{\em normalized} Gaussian sum-rules (NGSRs)~\cite{harnett_quark}
\begin{gather}
  N^{\qcd}_k (\hat{s}, \tau, s_0) =
  \frac{G^{\qcd}_{k} (\hat s, \tau, s_0) + \delta_{k,\,-1}\frac{1}{\sqrt{4\pi\tau}}
  \exp\left(\frac{-\hat{s}^2}{4\tau}\right)
  \Pi(0) }{M_{k,0} (\tau, s_0)+\delta_{k,\,-1}\Pi(0)}
\label{tom_norm_srk}
\\
  M_{k,n}(\tau, s_0)
 =\int\limits_{-\infty}^\infty 
\hat s^nG^\qcd_k (\hat s,\tau, s_0)\;\dif{\hat{s}}
\label{moments}
\end{gather}
which are related to the hadronic spectral function via
\begin{equation}\label{ngsr}
   N_k^{\qcd}(\hat{s},\tau,s_0) = \frac{ \frac{1}{\sqrt{4\pi\tau}} \int_{t_0}^{\infty} t^k
   \exp\left[\frac{-(\hat{s}-t)^2}{4\tau} \right] \rho^{{\rm had}}(t)\;\dif{t}}{\int_{t_0}^{\infty} t^k
   \rho^{{\rm had}}(t)\;\dif{t}} ~.
\end{equation}

For diagonal correlation functions the spectral function
obeys a positivity constraint so the NGSR must exist. For
non-diagonal correlators the possibility of state mixing
implies that $\rho^{\rm had}(t)$ could change sign, so it is
possible that either $M_{k,0} (\tau, s_0)$ or the
denominator on the right-hand sides of~\eqref{tom_norm_srk} or \eqref{ngsr} could be
zero.  In such situations, the GSRs~would have to be analyzed instead of the NGSRs.

We next consider the currents that will be used to probe the
gluonic and $q\bar q$ aspects of the scalar hadronic states.  
Refs.~\cite{narison,pascual} argue eloquently that the 
mixing of $q\bar q$ mesons and gluonium is unavoidable
because of the trace anomaly for the energy-momentum tensor $T_{\mu\nu}$~\cite{trace} 
\begin{equation}
T^\mu_\mu= \frac{1}{4}\beta(\alpha)G^a_{\mu\nu}G^{a\,\mu\nu}+\left[1+\gamma(\alpha)\right]\sum_f m_f\bar \psi_f\psi_f
\label{trace_anomaly}
\end{equation}
where 
\begin{gather}
  2\pi \alpha\beta\left(\alpha\right) =\nu^2\frac{\mathrm{d}}{\mathrm{d}\nu^2}\left(\frac{\alpha}{\pi}\right)=
  -\beta_0\left(\frac{\alpha}{\pi}\right)^2-\beta_1\left(\frac{\alpha}{\pi}\right)^3+\ldots
\\
  \beta_0 = \frac{11}{4}-\frac{1}{6} n_f~ ,~
  \beta_1=\frac{51}{8}-\frac{19}{24}n_f
\\
-2m\gamma(\alpha)=\nu^2\frac{\mathrm{d m}}{\mathrm{d}\nu^2}~.
\end{gather}
Equation~\eqref{trace_anomaly} actually contains \emph{two} multiplicatively-renormalizable 
(renormalization-group invariant) composite operators: $m\bar \psi \psi$ and 
$\beta G^2+4\gamma m\bar \psi\psi$. 
From a strictly field theoretical perspective, both are suitable choices for currents.  
However, the gluonic and/or $q\bar q$ nature of states which
couple to the current $\beta G^2+4\gamma m\bar \psi\psi$ would be
difficult to disentangle.
As such, we instead follow \cite{paver} and use renormalized currents 
\begin{gather}
 J_g=\alpha G^2~,~G_R^2=  \left(1+\frac{\beta_0}{\epsilon}\frac{\alpha}{\pi} \right) G^2_B
-4\frac{\alpha}{\pi}\frac{1}{\epsilon}\left(m_u\overline{u}u + m_d\overline{d}d\right)_B
+\ldots
\label{J_g}
\\
J_q= m_q\left(\overline{u}u + \overline{d}d\right)~,~m_q = \frac{1}{2}\left(m_u +m_d\right)~.
\label{J_q}
\end{gather}
where $R$ denotes a renormalized composite
operator and $B$ denotes  bare
quantities. 
Our convention for dimensional regularization uses $D=4+2\epsilon$ spacetime dimensions. 
Of course the form of the renormalized operator~\eqref{J_g}
necessarily underlies the renormalization-group invariance
of the trace anomaly~\eqref{trace_anomaly} (see 
{\it e.g.~}\cite{tarrach}).
However, the advantage of the current $J_g$ is that its
tree-level expansion is purely gluonic allowing a
qualitative separation of gluonic and $q\bar q$ degrees of
freedom.  Note that the use of a scalar tri-gluonium current, with three factors of the field strength rather than two as in \eqref{J_g}, does not seem to couple to the lightest state and mixes weakly \cite{paban}. 
The non-strange current $J_q$ has isospin $I=0$ and is renormalization-group invariant.

We define diagonal correlators corresponding to~(\ref{J_g}) and~(\ref{J_q})
as follows:
\begin{gather}
  \Pi_{gg}\left(Q^2\right)=\ii\int\,\mathrm{d}^4x\,\e^{\ii q\cdot x}\left\langle 0
  \vert T\left[ J_g(x)J_g(0)\right] \vert 0 \right\rangle
  \label{glue_corr}
\\
  \Pi_{qq}\left(Q^2\right) = \ii\int\,\mathrm{d}^4x\;\e^{\ii q\cdot x}
  \left\langle 0 | T \left[J_q (x) J_q (0)\right] |0\right\rangle
  \label{quark_corr}
\end{gather}
where $Q^2\equiv -q^2$.
Although both correlation functions are probes of scalar mesons, those states which have a
more significant overlap with the gluonic current should
predominate in \eqref{glue_corr};
those states which are dominantly of a non-strange
quark ($q\bar q$) nature should be more significant in
\eqref{quark_corr}.
A mixed state with substantial gluonic and quark components ({\it i.e.,} a state that
couples to both the gluonic and quark currents) should
self-consistently appear in an analysis of both correlation
functions.
In particular, independent predictions of identical-mass  states from
QCD sum-rule analyses of both~(\ref{glue_corr}) and~(\ref{quark_corr}) would be
indicative of mixing.  Note that the currents are simply probes of the actual hadronic spectrum and
do not impose a particular interpretation on the states: any state that has non-strange $\bar q q$ content would be probed by $J_q$, and any state with gluonic content would be probed by $J_g$ regardless of any additional content ({\it e.g.,} $s\bar s $) in the states.

In the scalar gluonic channel, a low-energy theorem (LET) \cite{let_glue}
\begin{equation}
  \Pi_{gg}(0)\equiv\lim_{Q^2\rightarrow 0} \Pi_{gg}(Q^2) = \frac{8\pi}{\beta_0} \langle \alpha G^2\rangle
\label{letDerek}
\end{equation}
allows construction of the $k=-1$ GSR.  The significance of instanton contributions in the overall consistency of the LET-sensitive
$k=-1$ sum-rule and the LET-insensitive $k\ge 0$ sum-rules was first  demonstrated for Laplace sum-rules \cite{DIL,inst_SR}.  A
similar consistency is observed for Gaussian sum-rules, but theoretical uncertainties are better controlled in the $k\ge 0$ GSRs
\cite{harnett_glue}; hence, we focus here on the $k=0$ GSRs for
both the diagonal gluonic and quark channels.
QCD expressions for the GSRs   $G_0^{(gg)}(\hat{s},\tau,s_0)$ and   $G_0^{(qq)}(\hat{s},\tau,s_0)$ corresponding to the diagonal correlation functions \eqref{glue_corr}, \eqref{quark_corr} 
can be found in \cite{harnett_glue,harnett_quark}.

\section{Non-Diagonal Correlation Function and GSRs of $q\bar q$ and Gluonic Currents}\label{corr_sec}
The 
non-diagonal correlation function for quark and gluonic currents
\begin{equation}
  \Pi_{gq}\left(Q^2\right)=\ii\int\,\mathrm{d}^4x\,\e^{\ii q\cdot x}\left\langle 0
  \vert T\left[ J_q(x)J_g(0)\right] \vert 0 \right\rangle
  ~,~ Q^2\equiv -q^2~,
  \label{mix_corr}
\end{equation}
contains perturbative, QCD condensate, and instanton contributions
\begin{equation}
  \Pi_{gq}\left(Q^2\right)=\Pi_{gq}^{pert}\left(Q^2\right)+\Pi_{gq}^{cond}\left(Q^2\right)
+\Pi_{gq}^{inst}\left(Q^2\right)~.
\end{equation}
  
\begin{figure}[hbt]
\centering
\includegraphics[scale=0.8]{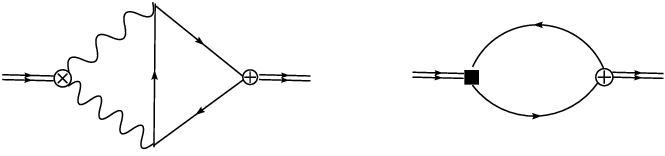}
\caption{
Leading order perturbative diagrams for the non-diagonal correlation function.  The symbol $\otimes$ denotes
the bare current $\alpha G_B^2$ within $J_g$  and $\oplus$
denotes the bare current $J_q$.  
In the second diagram, the solid square represents the $q\bar q$ term arising in the renormalization of $J_g$.
The Feynman diagrams were drawn with JaxoDraw \cite{jaxodraw}.
}
\label{pert_fig}
\end{figure}

The leading-order perturbative diagrams that contribute to $\Pi^{pert}_{gq}\left(Q^2\right)$ are given in Fig.~\ref{pert_fig}.  
The first diagram, a two-loop calculation, corresponds to the (bare) gluonic term in~\eqref{J_g}.
The second, a one-loop calculation, corresponds to the
(bare) quark term arising from composite-operator renormalization ({\it i.e.,} it is a renormalization-induced diagram).
Despite the differing number of loops, 
both diagrams are \order{\alpha}{2}.\footnote{Strange (and heavier) quarks originating
from composite operator renormalization in~\eqref{J_g} will
be suppressed by an additional factor of $\alpha^2$.}  
Also, both diagrams have the same \order{m_q}{2} leading
chiral behaviour because the quark loop in the first diagram provides a
$\mathcal{O}\left( m_q \right)$ 
chiral suppression factor.  The perturbative contributions can thus be separated into these bare and renormalization-induced 
diagrams
\begin{equation}
\Pi^{pert}_{gq}\left(Q^2\right)=\Pi^{bare}_{gq}\left(Q^2\right)+\Pi^{renorm}_{gq}\left(Q^2\right)~. 
\end{equation}

At leading chiral order in the $\overline{{\rm MS}}$ scheme, the result for  $\Pi^{bare}_{gq}$ 
(corresponding to the two-loop diagram in Fig.~\ref{pert_fig}) is 
\begin{equation}
\Pi^{bare}_{gq}\left(Q^2\right)
=\alpha^2m_q^2\left[\frac{3Q^2L}{\pi^3}\frac{1}{\epsilon}+\frac{3Q^2L}{\pi^3}\left(L-\frac{35}{6}\right)\right]
~,~ L=\log\left(\frac{Q^2}{\nu^2}\right)
\label{pert_diag}
\end{equation}
where $\nu$ is the renormalization scale.
We have ignored non-logarithmic terms in~\eqref{pert_diag} 
as they correspond to dispersion relation
subtraction constants which are eliminated upon forming the GSRs.  
The $\frac{L}{\epsilon}$ term in~\eqref{pert_diag} is problematic
since it cannot be renormalized away or absorbed into a dispersion-relation subtraction
constant.
However, the leading chiral order contribution $\Pi^{renorm}_{gq}$ arising from the
one-loop renormalization-induced diagram of Fig.~\ref{pert_fig}  is
\begin{equation}
\Pi^{renorm}_{gq}= - 3m_q^2Q^2 L\frac{\alpha^2}{\pi^3}
\frac{1}{\epsilon}
+3m_q^2Q^2\frac{\alpha^2}{\pi^3}\left( -\frac{L^2}{2} +2L\right)
\label{renorm_diag}
\end{equation}
where we have again ignored non-logarithmic terms.  An important, but subtle, aspect in the calculation of 
$\Pi^{renorm}_{gq}$ is retaining order $\epsilon$ contributions from the loop integrals that lead to finite terms when combined with the renormalization constant appearing in \eqref{J_g}.   This methodology is necessary and can be  verified for well-known correlation functions ({\it e.g.,} light-quark pseudoscalar currents).
The offending $\frac{L}{\epsilon}$ term  in \eqref{pert_diag} is thus cancelled by a compensating term in \eqref{renorm_diag}, and we are
left with the leading-order $\overline{{\rm MS}}$-scheme perturbative contribution to the non-diagonal correlation function:\footnote{
The numerical coefficients in this result disagree with those presented in Ref.\ \cite{paver}, although the overall chiral and logarithmic dependence is identical.  We have checked our calculational methodology by verifying the pseudoscalar results in \cite{paver} and believe that~\eqref{pert_result} is correct.
}
\begin{gather}
\Pi_{gq}^{pert}\left(Q^2\right)=m_q^2Q^2\left[A_0L+A_1L^2  \right]\\
A_0=-\frac{23}{2\pi}\left(\alpi\right)^2 ~,~A_1=\frac{3}{2\pi}\left(\alpi\right)^2~.
\label{pert_result}
\end{gather}

The QCD condensate contributions \cite{SVZ} to the non-diagonal correlator, including operators up  to dimension-five, can be written as an operator-product expansion (OPE)
\begin{gather}
\Pi_{gq}^{cond}\left(Q^2\right)= C_{QQ} \left\langle \bar{q}q\right\rangle
+  C_{GG} \left\langle \alpha G^2\right\rangle
+C_{QGQ} \left\langle \bar q \sigma G q\right\rangle
\label{corr_cond_new}
\\
\left\langle \bar{q}q\right\rangle = \frac{1}{2}\langle \bar u u+\bar d d\rangle
~,~ \left\langle \alpha G^2\right\rangle=\left\langle \alpha G^a_{\mu\nu}G^{a\,\mu\nu}\right\rangle
~,~ \left\langle \bar q \sigma G q\right\rangle= 
\left\langle \bar q g \frac{\lambda^a}{2} \sigma^{\mu\nu} G^a_{\mu\nu} q\right\rangle~.
\end{gather}
As will be shown below, the chiral violating effects of the quark condensate $\left\langle \bar{q}q\right\rangle$ 
and mixed condensate $\left\langle \bar q \sigma G q\right\rangle$ will dominate that of the gluon condensate 
$\left\langle \alpha G^2\right\rangle$.  
To obtain the leading chiral order behaviour of the OPE coefficients, it is necessary to include higher-order mass contributions that result in operator mixing \cite{jamin,bagan}. In particular, the naively-calculated unmixed OPE coefficients $E_{QQ}$, $E_{GG}$, and $E_{QGQ}$ are related to the coefficients in \eqref{corr_cond_new} by \cite{jamin}
\begin{gather}
 E_{QQ}=C_{QQ}~,~E_{QGQ}=C_{QGQ}
\\
C_{GG}=E_{GG}+\frac{1}{12\pi m_q}C_{QQ}-\frac{m_q}{2\pi} \log{\left(\frac{m_q^2}{\nu^2}\right)}C_{QGQ}~.
\label{mixing_eq}
\end{gather}
In principle, the coefficient of the identity operator ({\it i.e.,} perturbative contributions) also mix with the quark condensate coefficients, but such terms are proportional to $m_q^3 C_{QQ}$ and are therefore sub-leading in the quark mass compared with \eqref{pert_result}.

\begin{figure}[hbt]
\centering
\includegraphics[scale=0.8]{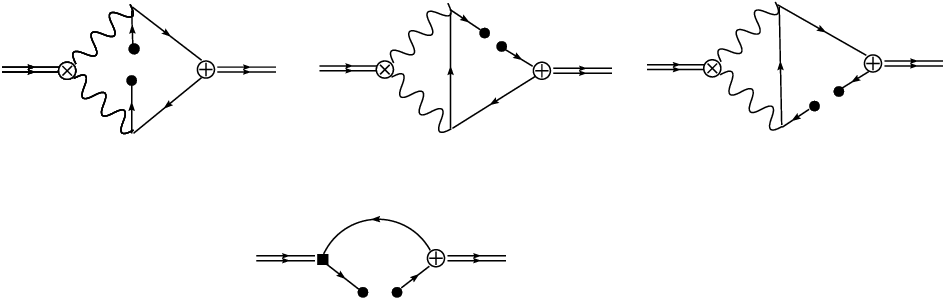}
\caption{
Leading $\alpha$ order quark condensate diagrams for the non-diagonal correlation function.  The solid circles on the
quark lines denote insertion of plane-wave states or coordinate space vacuum expectation values for evaluation of the operator-product expansion coefficients.  All other notations are identical to Fig.~\ref{pert_fig}.
}
\label{quark_cond_fig}
\end{figure}

The quark condensate contribution to the non-diagonal
correlator is easily calculated to leading chiral order using any of
the equivalent methods for evaluating OPE
coefficients~\cite{pwcs}.
As in the perturbative case, in principle there are two classes of
diagrams, both of which are depicted in Fig.~\ref{quark_cond_fig}. However, the renormalization-induced diagram of Fig.~\ref{quark_cond_fig}
is chirally-suppressed relative to~\eqref{qq_result} and so represents subleading effects.
Computing the first set  of diagrams in Fig.~\ref{quark_cond_fig} therefore gives the leading-order quark-condensate contributions 
\begin{equation}
C_{QQ}=-\frac{8}{\pi} \alpha^2m_q \log{\left(\frac{Q^2}{\nu^2}\right)}~.
\label{qq_result}
\end{equation}

\begin{figure}[hbt]
\centering
\includegraphics[scale=0.4]{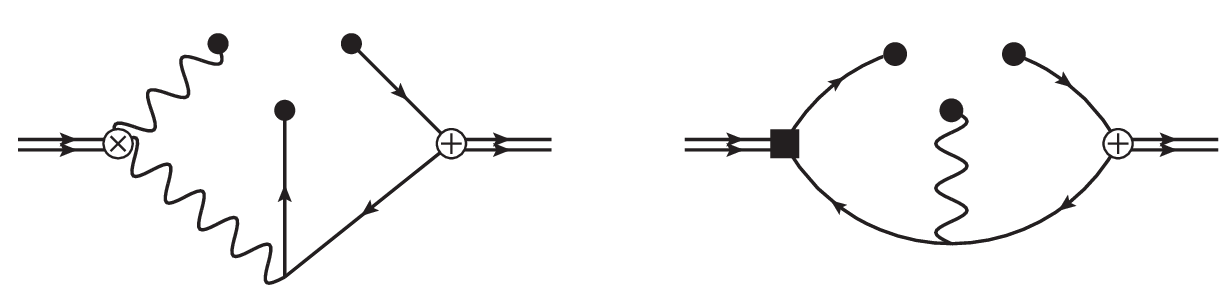}
\caption{
Leading order mixed condensate diagrams for the non-diagonal correlation function.  The solid circles on the
quark lines denote insertion of plane-wave states or coordinate space vacuum expectation values for evaluation of the operator-product expansion coefficients.  All other notations are identical to Fig.~\ref{pert_fig}.
}
\label{mixed_diag}
\end{figure}

Figure~\ref{mixed_diag} shows two of the diagrams that contribute to the  mixed condensate OPE coefficient.  Within fixed-point gauge methods,  the two-quark  vacuum expectation value appearing in  the renormalization-induced diagram 
of Fig.~\ref{quark_cond_fig} also introduces the  mixed-condensate (see {\it e.g.,} Ref.~\cite{elias}).  However, the 
two renormalization-induced diagrams are suppressed by a factor of $\alpha$ and additional factors of the quark mass
compared with the first diagram of Fig.~\ref{mixed_diag}, resulting in the following leading-order OPE coefficient
\begin{equation}
C_{QGQ}=\frac{4\alpha m_q}{Q^2}~.
\label{mix_coeff} 
\end{equation}

\begin{figure}[hbt]
\centering
\includegraphics[scale=0.7]{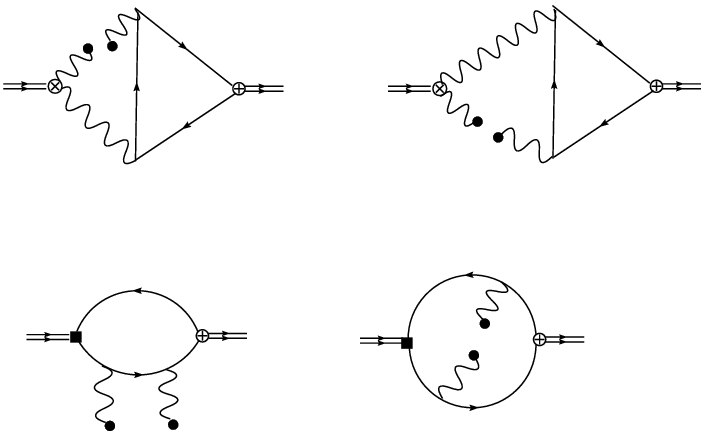}
\caption{
Leading order gluon condensate diagrams for the non-diagonal correlation function.  The solid circles on the
gluon lines denote insertion of plane-wave states  for evaluation of the operator-product expansion coefficients.  All other notations are identical to Fig.~\ref{pert_fig}.
}
\label{glue_cond_fig}
\end{figure}

In the absence of operator mixing, the  diagrams that could lead to the gluon condensate OPE coefficient $E_{GG}$ are shown in Fig.~\ref{glue_cond_fig}.  However, the  renormalization-induced diagrams are higher-order in $\alpha$ and hence are subleading.  Because of infrared divergences, it is necessary to retain the quark mass until the last steps of the calculation and then extract the leading chiral behaviour.  Using plane-wave methods with $m_u=m_d=m$ we find 
\begin{equation} 
E_{GG}=2i\frac{32\pi\alpha m_q^2 }{D^2 q^2}  
\left[
m^2(8-4D)I_1-2D I_2+(D^2-4D+8) I_3\right]~,
\end{equation}
where 
\begin{gather}
I_1=\frac{1}{\nu^{2\epsilon}}\int \frac{\mathrm{d}^Dk}{\left(2\pi\right)^D}\frac{1}{\left(k^2-m^2\right)^2\left[ (q-k)^2-m^2\right]}
=\frac{\ii}{16\pi^2}\frac{1}{q^2\sqrt{1-4m^2/q^2}}\log{\left[\frac{\sqrt{1-4m^2/q^2}+1}{\sqrt{1-4m^2/q^2}-1}\right]}~,
\\
I_2=\frac{1}{\nu^{2\epsilon}}\int \frac{\mathrm{d}^Dk}{\left(2\pi\right)^D}\frac{1}{\left(k^2-m^2\right)^2}
=\frac{\ii}{16\pi^2}\left(\frac{m^2}{4\pi\nu^2}\right)^\epsilon
\Gamma(-\epsilon)~,
\\
\begin{split}
I_3&=\frac{1}{\nu^{2\epsilon}}\int \frac{\mathrm{d}^Dk}{\left(2\pi\right)^D}\frac{1}{\left(k^2-m^2\right)\left[ (q-k)^2-m^2\right]}
\\
&=\frac{\ii}{16\pi^2}\left(-\frac{1}{\epsilon}-\gamma-\log{\left(\frac{m^2}{4\pi\nu^2}\right)}
+2-\sqrt{1-4m^2/q^2}
\log{\left[\frac{\sqrt{1-4m^2/q^2}+1}{\sqrt{1-4m^2/q^2}-1}\right]}
\right)~.
\end{split}
\end{gather}
The divergences in $I_3$ and $I_2$ cancel, leaving a finite result as required given  the $m_q^2/q^2$ pre-factor.  The logarithmic correction from $I_1$ is seen to be subleading compared with the logarithmic correction from $I_3$.  Thus the leading-chiral gluon condensate contribution to the non-diagonal unmixed OPE coefficient is 
\begin{equation}
E_{GG}=\frac{2\alpha m_q^2}{\pi Q^2} \left[3-\log{\left(\frac{Q^2}{m^2}\right)}  \right]~.
\label{EGG_result}
\end{equation}
The $\log{\left(m^2\right)}$ infrared divergence in \eqref{EGG_result} is now cancelled when \eqref{EGG_result} and \eqref{mix_coeff}
are substituted in \eqref{mixing_eq}, resulting in the leading-order contribution to the gluon condensate OPE coefficient
\begin{equation}
C_{GG}=\frac{2\alpha m_q^2}{\pi Q^2}\left[3-\log{\left(\frac{Q^2}{\nu^2}\right)}  \right]~.
\label{GG_result} 
\end{equation}
The cancellation of the infrared divergence follows from the methodology of \cite{jamin}, and provides a consistency check on our calculation of the OPE coefficients.

Finally, the result of our calculation for the  single instanton \cite{basic_instanton} contributions 
 ({\it i.e.,} multi-instanton effects are negligible~\cite{schaefer_shuryak}) 
 to the non-diagonal correlator in the dilute instanton liquid model \cite{DIL} are:
\begin{equation}
\Pi^{inst}_{gq}\left(Q^2\right)= -8\sqrt{3 n_c} m_q\rho Q^2 \sqrt{\rho^2Q^2} K_1\left(\sqrt{\rho^2Q^2}\right) K_2\left(\sqrt{\rho^2Q^2}\right) ~,
\label{inst_result}
\end{equation}
where $K_n$ is a modified Bessel function in the conventions of \cite{abr}.

Combining Eqs.~\eqref{pert_result}, \eqref{qq_result}, \eqref{mix_coeff}, \eqref{GG_result}, and \eqref{inst_result}, we have the
leading-order  chiral and $\alpha$ contributions to the non-diagonal correlation function of gluonic and (non-strange)
$I=0$ quark currents in the $\overline{\rm MS}$ scheme:
\begin{gather}
\begin{split}
\Pi_{gq}\left(Q^2\right)=&m_q^2Q^2\left[A_0L+A_1L^2  \right]
\\
&+ m_q\langle \bar qq\rangle C_0 L 
+m_q^2\langle \alpha G^2\rangle \frac{1}{Q^2}\left[B_0+B_1 L 
\right]
+m_q\left\langle \bar q \sigma G q\right\rangle\frac{D_0}{Q^2}
\\&-
8\sqrt{3 n_c} m_q \rho Q^2 \sqrt{\rho^2Q^2} K_1\left(\sqrt{\rho^2Q^2}\right) K_2\left(\sqrt{\rho^2Q^2}\right)
\end{split}
\label{final_corr}
\\
C_0=-8\pi\left(\frac{\alpha}{\pi}\right)^2~, ~D_0=4\alpha
\\
B_0=6\frac{\alpha}{\pi}~,~B_1=-2\frac{\alpha}{\pi}~,
\end{gather}
with $A_0$, $A_1$  given in~\eqref{pert_result}.
From the correlation function, the $k=0$ GSR can be calculated as outlined in Section \ref{gsr_sec} and Ref.~\cite{harnett_glue}:
\begin{equation}
\begin{split}
G^{(gq)}_0\left(\hat s,\tau,s_0\right)=&\frac{m_q^2}{\sqrt{4\pi\tau}}\int\limits_0^{s_0}\dif{t}\, 
\exp{\left[-\frac{\left(t-\hat s\right)^2}{4\tau}\right]} t\left( A_0+2A_1\log\left[\frac{t}{\sqrt{\tau}}\right]\right)+B_0m_q^2\langle \alpha G^2\rangle\frac{1}{\sqrt{4\pi\tau}}\exp{\left(-\frac{\hat s^2}{4\tau}\right)}
\\
&+B_1m_q^2\langle \alpha G^2\rangle\frac{1}{\sqrt{4\pi\tau}}
\lim_{\eta\to 0}\left\{
\int\limits_\eta^{s_0}\dif{t}\,\frac{1}{t}\exp{\left[-\frac{\left(t-\hat s\right)^2}{4\tau}\right]}
+\log{\left(\frac{\eta}{\sqrt{\tau}}\right)}\exp{\left(-\frac{\hat s^2}{4\tau}\right)}
\right\}
\\
&-C_0m_q\langle\bar qq\rangle\frac{1}{\sqrt{4\pi\tau}}\int\limits_0^{s_0}\dif{t}\,
\exp{\left[-\frac{\left(t-\hat s\right)^2}{4\tau}\right]}
+D_0 m_q\left\langle\bar q\sigma G q\right\rangle \frac{1}{\sqrt{4\pi\tau}}\exp{\left(-\frac{\hat s^2}{4\tau}\right)}
\\
&+2\sqrt{3n_c}\pi m_q\rho^2 
\frac{1}{\sqrt{4\pi\tau}}
\int\limits_0^{s_0}\dif{t}\, \exp{\left[-\frac{\left(t-\hat s\right)^2}{4\tau}\right]}
t\sqrt{t}\left[
J_1\left(\rho\sqrt{t}\right)Y_2\left(\rho\sqrt{t}\right)+J_2\left(\rho\sqrt{t}\right)Y_1\left(\rho\sqrt{t}\right)
\right]
\end{split}
\label{gq_GSR}
\end{equation}
where in practice, the limit $\eta\to 0$ is implemented numerically with values of $\eta<10^{-4}\,{\rm GeV^2}$.   
Due to renormalization-group scaling of the GSRs~\cite{harnett_quark,gauss}, we have set $\nu^2=\sqrt{\tau}$ in \eqref{gq_GSR}, and 
hence $m_q$ and $\alpha$ are implicitly the leading-order versions of the running quantities
 evaluated at the scale $\nu^2=\sqrt{\tau}$ for 
three active flavours in the $\overline{\rm{MS}}$ scheme
\begin{equation}
\begin{gathered}
  \frac{\alpha (\nu^2)}{\pi} = \frac{1}{\beta_0 L}~,~
m_q\left(\nu^2\right)=  \frac{\hat m_q}{\left(\frac{1}{2}L\right)^{\frac{4}{9}}}
\\
  L=\log\left(\frac{\nu^2}{\Lambda^2}\right)~,~
  \beta_0=\frac{9}{4}
\end{gathered}
\label{run_coupling_mass}
\end{equation}
where $\hat m_q$ is the renormalization-group invariant quark mass parameter and 
 $\Lambda_{\overline{MS}}\approx 300\,{\rm MeV}$
consistent with current estimates of $\alpha(M_\tau)$~\cite{pdg}.
The normalized GSR $N^{(gq)}_0\left(\hat s,\tau,s_0\right)$ associated with \eqref{gq_GSR} is defined by \eqref{tom_norm_srk}:
\begin{equation}
 N^{(gq)}_0\left(\hat s,\tau,s_0\right)=\frac{G^{(gq)}_0\left(\hat s,\tau,s_0\right)}{M^{(gq)}_{0,0}\left(\tau,s_0\right)}
~,~ M^{(gq)}_{0,0}(\tau, s_0)
 =\int\limits_{-\infty}^\infty 
G^{(gq)}_0 (\hat s,\tau, s_0)\;\dif{\hat{s}} ~.
\label{N_gq}
\end{equation}

Since a low-energy theorem exists for the non-diagonal correlator \cite{let_glue}
\begin{equation}
 \Pi_{gq}(0)=\frac{48\pi}{9}m_q\langle\bar q q\rangle~,
\end{equation}
the LET-sensitive $k=-1$ GSR is also relevant. Again using the methods outlined in Section \ref{gsr_sec} and Ref.~\cite{harnett_glue}, the corresponding results for the $k=-1$ GSR are 
\begin{equation}
\begin{split}
G^{(gq)}_{-1}\left(\hat s,\tau,s_0\right)=&\frac{m_q^2}{\sqrt{4\pi\tau}}\int\limits_0^{s_0}\dif{t}\, 
\exp{\left[-\frac{\left(t-\hat s\right)^2}{4\tau}\right]} \left( A_0+2A_1\log\left[\frac{t}{\sqrt{\tau}}\right]\right)
\\
&-C_0m_q\langle\bar qq\rangle\frac{1}{\sqrt{4\pi\tau}}\lim_{\eta\to 0}\left\{
\int\limits_\eta^{s_0}\dif{t}\,\frac{1}{t}\exp{\left[-\frac{\left(t-\hat s\right)^2}{4\tau}\right]}
+\log{\left(\frac{\eta}{\sqrt{\tau}}\right)}\exp{\left(-\frac{\hat s^2}{4\tau}\right)}
\right\}
\\
&+B_1m_q^2\langle \alpha G^2\rangle\frac{1}{\sqrt{4\pi\tau}}
\lim_{\eta\to 0}\left\{
\int\limits_\eta^{s_0}\dif{t}\,\frac{1}{t^2}\exp{\left[-\frac{\left(t-\hat s\right)^2}{4\tau}\right]}
-\frac{1}{\eta}\exp{\left(-\frac{\hat s^2}{4\tau}\right)}
\right\}
\\
&+B_0m_q^2\langle \alpha G^2\rangle\frac{1}{\sqrt{4\pi\tau}}\frac{\hat s}{2\tau}\exp{\left(-\frac{\hat s^2}{4\tau}\right)}+D_0 m_q\left\langle\bar q\sigma G q\right\rangle \frac{1}{\sqrt{4\pi\tau}}\frac{\hat s}{2\tau}\exp{\left(-\frac{\hat s^2}{4\tau}\right)}
\\
&+2\sqrt{3n_c}\pi m_q \rho^2 
\frac{1}{\sqrt{4\pi\tau}}
\int\limits_0^{s_0}\dif{t}\, \exp{\left[-\frac{\left(t-\hat s\right)^2}{4\tau}\right]}
\sqrt{t}\left[
J_1\left(\rho\sqrt{t}\right)Y_2\left(\rho\sqrt{t}\right)+J_2\left(\rho\sqrt{t}\right)Y_1\left(\rho\sqrt{t}\right)
\right]
\\
&+\frac{1}{\sqrt{4\pi\tau}}\exp{\left(-\frac{\hat s^2}{4\tau}\right)}\frac{16\sqrt{3n_c} m_q}{\rho}~.
\end{split}
\label{gq_m1_GSR}
\end{equation}
Note that the last term in~\eqref{gq_m1_GSR} has a functional dependence identical to the LET term in~\eqref{tom_norm_srk}, and hence there is an LET-like instanton contribution  for the non-diagonal GSR similar to the diagonal gluonic case \cite{harnett_glue}.  
The NGSR $N^{(gq)}_{-1}\left(\hat s,\tau,s_0\right)$ associated with \eqref{gq_m1_GSR} is defined by \eqref{tom_norm_srk}:
\begin{equation}
 N^{(gq)}_{-1}\left(\hat s,\tau,s_0\right)=
\frac{G^{(gq)}_{-1}\left(\hat s,\tau,s_0\right)+\frac{1}{\sqrt{4\pi\tau}}
  \exp\left(\frac{-\hat{s}^2}{4\tau}\right)
  \Pi_{gq}(0)}{M^{(gq)}_{-1,0}\left(\tau,s_0\right)+\Pi_{gq}(0)}
~,~M^{(gq)}_{-1,0}(\tau, s_0)
 =\int\limits_{-\infty}^\infty 
G^{(gq)}_{-1} (\hat s,\tau, s_0)\;\dif{\hat{s}} ~.
\end{equation}

The QCD input parameters appearing within the non-diagonal NGSRs will now be specified.  For the gluon condensate we employ the (central) value from \cite{nar97}
\begin{equation}\label{dimfour}
 \langle\alpha G^2\rangle
      = (0.07\pm 0.01)\, {\rm GeV^4} ~,
\end{equation}
and the quark condensate is determined by the PCAC relation
\begin{equation}
 m_q\left\langle\bar q q\right\rangle=-\frac{1}{2}f_\pi^2m_\pi^2~,~f_\pi=93\,{\rm MeV}~.
\label{quark_cond}
\end{equation}
For our purpose, the expression of the mixed condensate in terms of the quark condensate \cite{mixed_cond_ref} is the most useful:
\begin{equation}
 \left\langle \bar q \sigma Gq\right\rangle=M_0^2\left\langle\bar q q\right\rangle~,~M_0^2=\left(0.8\pm 0.1\right) \,{\rm GeV^2}~.
\label{mixed_cond_value}
\end{equation} 
In addition, the dilute instanton liquid (DIL) model parameters (which have an estimated uncertainty of about 15\%)~\cite{DIL}
\begin{equation} \label{DILparams}
  n_{{c}} = 8.0\times 10^{-4}\ {\rm GeV^4}~,~\rho =
  \frac{1}{0.6}\ {\rm GeV}^{-1}~
\end{equation}
will be employed.  
The NGSRs for the diagonal correlators  do not require knowledge of the quark masses; in the diagonal gluonic case this occurs because the leading chiral behaviour is independent of the quark masses, while the 
diagonal $q\bar q$ case is proportional to $m_q^2$ and hence the quark mass dependence cancels when forming the NGSR.  However, the non-diagonal case has terms of differing chiral order and therefore requires input of the quark mass.  Unfortunately, $m_q$ is not known very accurately; we will use the Particle Data Group range for the $2\,{\rm GeV}$ $\overline{{\rm MS}}$ mass~\cite{pdg}:
\begin{equation}
2.5\,{\rm MeV} <m_q(2\,{\rm GeV})<5.5\,{\rm MeV}~.
\label{quark_mass}
\end{equation}
The implications of the large uncertainty in $m_q$ within our analysis will be discussed in more detail below.

The non-diagonal correlator and its associated GSR have quite distinct chiral behaviour compared with the diagonal correlators.  In the diagonal case, the perturbative, condensate, and instanton corrections all appear with identical powers of the quark mass.  However, in the non-diagonal case the perturbative and gluon condensate corrections are chirally-suppressed compared with the quark condensate, mixed condensate, and instanton terms.  
One can understand this chiral behaviour in the non-diagonal case by recognizing that the operator 
$\bar u(x)u(x)+\bar d(x)d(x)$ appearing in $J_q$ violates chiral symmetry, and hence the 
chiral-conserving gluon condensate (and perturbative corrections) must have an additional chiral suppression because the gluon condensate would be non-zero in the limit of vanishing quark mass.  Similarly, the chiral-violating condensates (and instanton) do not require an additional mass chiral suppression because chiral symmetry is restored when the condensates are zero.
The implications of the quark mass suppression  of the chiral-preserving terms, and the comparative enhancement of chiral-violating effects, 
will be discussed below.

\section{Analysis of Gaussian Sum-Rules for $q\bar q$ and Gluonic Currents}\label{analysis_sec}
The general strategy for analysis of NGSRs involves matching the QCD expression with a parametrized model for $\rho^{\rm had}(t)$ in \eqref{ngsr}.
Correlation functions of vector and axial-vector $q\bar q$ currents  can be directly related to experimental data ({\it e.g.,} $R(s)$, $\tau$ decays), but in the case of gluonium there is no direct connection with experimental observables. 
The narrow resonance approximation is the most common choice made for Laplace sum-rule analyses of gluonium, with either a single (narrow) resonance \cite{kisslinger,NSVZ_glue,kisslinger_inst,lsr_narrow} to examine the dominant gluonic state or two (narrow) resonances \cite{narison,ailin,bagan_glue} to explore the possibility of  $q\bar q$-gluonium mixtures.  Laplace sum-rule gluonium analyses which go beyond the narrow width approximation include a single Breit-Wigner resonance skewed by kinematic factors~\cite{liu}, and an interpolation between the LET and continuum behaviour \cite{pascual}.  Finite-energy sum-rule analyses of scalar gluonium include narrow resonance models \cite{krasnikov} and incorporate resonance widths through step functions \cite{shifman} and  
Breit-Wigner resonances \cite{dominguez} with kinematic skewing.\footnote{Ref.~\cite{dominguez} also uses the Gaussian sum-rule diffusion equation analysis to constrain the QCD continuum. As discussed in Section~\ref{gsr_sec}, our approach based on NGSRs provides information that is {\em independent} of the FESR duality constraint.}

GSR analyses of gluonium have employed single and double narrow resonance models in addition to a variety of models that incorporate resonance widths \cite{harnett_glue, harnett_proc}. However,  inclusion of width effects do not lead to appreciable improvement in the agreement between the QCD expression and phenomenological model.  We attribute this to the large value of the QCD-limited width of the Gaussian kernel $2\sqrt{\tau}\ge 2\,{\rm GeV^2}$ in \eqref{basic_gauss}  which obscures resonance-width effects.  As in our previous combined analysis of the diagonal quark and gluonic GSRs \cite{harnett_proc}, we thus choose a double narrow resonance model for capturing the essential features of the analysis.

In general, the analysis of NGSRs in the double narrow resonance model has the form
 \begin{gather} 
\frac{1}{\pi}\rho^{\text{had}}(t) =  f_1^2 \delta(t-m_1^2) + f_2^2 \delta(t-m_2^2)~,
\\
  N^{\qcd}_0(\hat{s},\tau,s_0)  = \frac{1}{\sqrt{4\ppi\tau}}
  \left\{r_1 \exp\left[-\frac{(\hat{s}-m_1^2)^2}{4\tau}\right]
  +       r_2 \exp\left[-\frac{(\hat{s}-m_2^2)^2}{4\tau}\right]
  \right\}~,
\label{phenom_double}
\\
r_1=\frac{f_1^2}{f_1^2+f_2^2}~,~r_2=\frac{f_2^2}{f_1^2+f_2^2}~,~r_1+r_2=1~
\label{r_def}
\end{gather}
where $f_1$, $f_2$ denote the couplings of the resonances to the currents under consideration and $m_1<m_2$.  As outlined in
Refs.~\cite{harnett_glue, harnett_proc}. 
The GSR moments  \eqref{moments} are the most useful quantities for extracting the resonance parameters from the QCD expression.  In particular, the 
first-order moments provide a measure of the peak of the GSR
\begin{equation}
P\left(\tau,s_0\right)=\frac{M_{0,1}\left(\tau,s_0\right)}{M_{0,0}\left(\tau,s_0\right)}~,
\label{first_moment}
\end{equation}
second-order moments provide a measure of the GSR width 
\begin{equation}
  \sigma^2\left(\tau,s_0\right)=
 \frac{M_{0,2}\left(\tau,s_0\right)}{M_{0,0}\left(\tau,s_0\right)}
-\left[P\left(\tau,s_0\right)\right]^2~,
\label{sigma_combo}
\end{equation}
and the third-order moments provide a measure of the GSR asymmetry
\begin{equation}
A\left(\tau,s_0\right)
=\frac{M_{0,3}\left(\tau,s_0\right)}{M_{0,0}\left(\tau,s_0\right)}
-3\sigma^2\left(\tau,s_0\right)P\left(\tau,s_0\right) -\left[P\left(\tau,s_0\right)\right]^3~.
\label{dist_asymm}
\end{equation}
The double-resonance phenomenological parameters defined by
\begin{equation}
r=r_1-r_2\,,~y=m_1^2-m_2^2\,,~z=m_1^2+m_2^2~,
\end{equation}
are then related to the moments by
\begin{gather}
 z  =   2P + \frac{A}{\sigma^2-2\tau}
\label{z_moms}\\
 y  =   \frac{ -\sqrt{A^2 + 4(\sigma^2-2\tau)^3}}{\sigma^2-2\tau}
\label{y_moms}\\
 r  =   \frac{A}{\sqrt{A^2 + 4(\sigma^2-2\tau)^3}}~,
\label{r_moms}
\end{gather}
where the $\tau$, $s_0$ dependence of the moments has been suppressed for brevity.
The quantity $\sigma^2-2\tau$ appearing in Eqs.~\eqref{z_moms}--\eqref{r_moms} is particularly important because it is 
a clear diagnostic of multiple resonances.  This can be seen from 
\begin{equation}
\sigma^2-2\tau=\frac{1}{4}y^2\left(1-r^2\right)>0~,
\label{sig_rel}
\end{equation}
which indicates that a second resonance cannot be absorbed into the continuum if the QCD value of the second-order GSR moments exceed the natural Gaussian width of $2\tau$.  

In general, the resonance parameters depend on $\tau$ and $s_0$ through the QCD values of the moments. 
Apart from the previously-discussed QCD constraints,  
 $\tau$ is a free parameter and therefore the resonance parameters should be largely independent of $\tau$;  
residual $\tau$ dependence can be interpreted as a source of theoretical uncertainty.  However, the continuum threshold $s_0$ appears within the QCD expression, so a criterion must be established for optimizing $s_0$.
Various approaches to this optimization will be discussed below.

\begin{figure}[htb]
\centering
\includegraphics[scale=0.4]{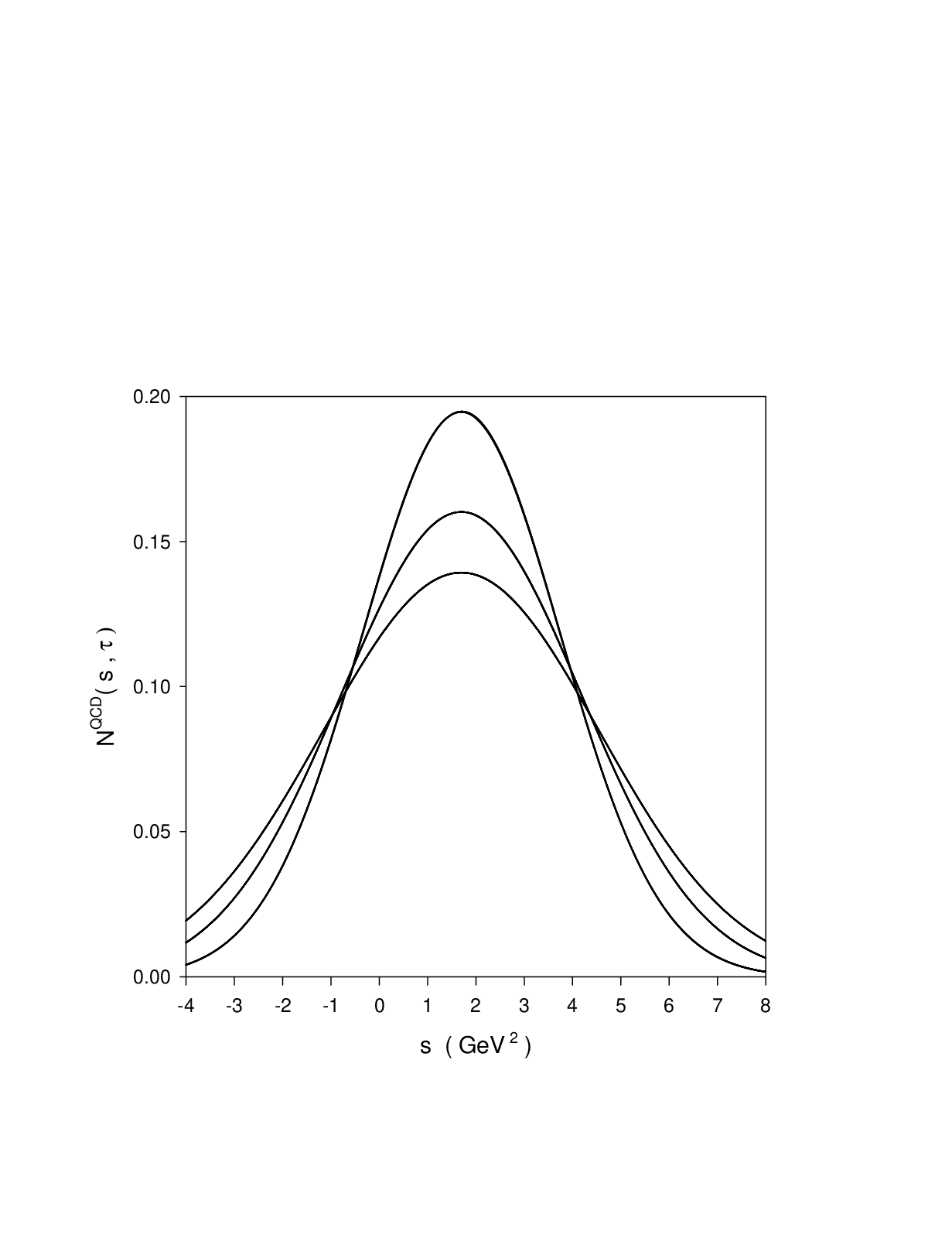}
\caption{
Comparison of the
QCD theoretical expression for  $N_0^{(gg)}\left(\hat s, \tau,s_0\right)$ with the
double narrow resonance phenomenological model in the diagonal gluonic case.
The $\tau$ values used for the three pairs of curves, from top to bottom in the figure, are respectively
 $\tau=2.0\,{\rm GeV}^4$, $\tau=3.0\, {\rm GeV}^4$, and $\tau=4.0\,
{\rm GeV}^4$.
Note the almost perfect overlap between the theoretical expression and the
phenomenological models.
A qualitatively similar agreement between the
 double narrow resonance model and the QCD expression exists for the  diagonal quark NGSR.
}
\label{twores_fig}
\end{figure}

In the case of the diagonal quark and gluonic cases, $s_0$ was constrained by 
studying the $\tau$ dependence of the value $\hat s_{peak}$ at which the QCD expression for the NGSR reaches its maximum value.   This $\tau$ dependence is then compared with that arising from a double-resonance model, and $s_0$ is constrained by optimizing the agreement between them.  This procedure for optimizing $s_0$ and determining the resonance parameters has been confirmed by a more numerically-intensive
multi-parameter fit of $s_0$ and the resonance parameters \cite{harnett_quark,ps_glue}.  
The  resonance parameters resulting from our previous analyses  of the diagonal gluonic and diagonal quark NGSRs
\cite{harnett_glue,harnett_quark,harnett_proc} are summarized in Table \ref{doubres_tab} (the details of the non-diagonal case will be discussed below).  
The double narrow resonance model results in excellent agreement with the QCD expression as illustrated in Figure \ref{twores_fig} \cite{harnett_glue};
there is no indication of discrepancies  that would require a more elaborate phenomenological model ({\it e.g.,} additional states, resonance widths).
 For the diagonal NGSRs, the uncertainties associated with the QCD input parameters have been found to be 10\%  for $\{r^{(gg)}_2,~r^{(gg)}_1\}$, 
and at most $0.2\,{\rm GeV}$ for the masses with  a correlated  effect  that leads to a relatively stable  mass splitting $m_2-m_1\approx 0.4\,{\rm GeV}$ \cite{harnett_glue}.

\begin{table}[hbt]
  \centering
  \begin{tabular}{||c|c|c|c|c|c||}
    \hline\hline
Sum-Rule &  $m_1$ (GeV) & $m_2$ (GeV)  & $r_1$ & $r_2$ & $s_0$ (${\rm GeV^2}$)  \\
    \hline\hline
diagonal: gluonic-gluonic   &  $0.98$ & $1.4$ & $0.28$ & $0.72$ & $2.30$ \\\hline
diagonal: $\bar q q$-$\bar q q$   & $0.97$ & $1.4$ & $0.63$ & $0.37$ & $2.60$\\\hline
non-diagonal: gluonic-$\bar q q$ & $0.84$ & $1.4$ & $0.44$ & $0.56$ & $2.75$\\
    \hline\hline
  \end{tabular}
 \caption{Analysis results from the diagonal and non-diagonal NGSRs of gluonic and $\bar q q$ currents  in the double narrow resonance model.  Central values of the QCD input parameters have been employed.}
\label{doubres_tab}
\end{table}

The remarkable agreement between the resonance masses resulting from independent analysis of the diagonal gluonic and diagonal $q\bar q$ NGSRs suggests the existence of states with masses of approximately $1.0\,{\rm GeV}$ and $1.4\,{\rm GeV}$  that couple to mixtures of gluonium and $q\bar q$ currents, with the heavier state being slightly more gluonic because of its stronger coupling to gluonic currents and  weaker coupling to $q\bar q$ currents.  
This consistency of the mass predictions in the two channels is precisely what is
expected for hadronic states that couple to both  gluonium and quark currents.
The results also indicate that the mixing is rather strong (consistent with the conclusions of \cite{narison_review,narison}) because $r_1$ and $r_2$ are not appreciably different, and hence the non-diagonal correlator must also contain clear signals of this strong mixing to validate this scenario.
In other words, a definitive signal of states that are $q\bar q$-gluonic mixtures is their consistent appearance with the same mass in {\em all three} cases 
(diagonal gluonic, diagonal quark, and non-diagonal gluonic-quark)
since such mixtures would necessarily couple to both the gluonic and $q\bar q$ currents.


Before proceeding with a detailed analysis of the non-diagonal correlator, we consider the approximate scales associated with the couplings of the resonances to the gluonic and scalar currents.   The  perturbative corrections in the diagonal correlators (see Refs.~\cite{harnett_glue,harnett_quark})
  imply that
\begin{equation}
f^2_g\sim \left(\frac{\alpha}{\pi}\right)^2 E^4~,~f^2_q\sim m_q^2 E^2~,
\label{diag_scales}
\end{equation}
where $f_g$ and $f_q$ respectively denote the resonance couplings to the gluonic and $q\bar q$ currents and $E$ is a characteristic sum-rule energy scale of order $E\sim 1\,{\rm GeV}$.  In the simplest single-angle mixing scenario, the non-diagonal correlator will be proportional to $f_gf_q\sin{2\theta}$ where $\theta$ is the mixing angle. The perturbative corrections to the non-diagonal correlator~\eqref{final_corr} then imply
\begin{equation}
 f_gf_q\sin{2\theta}\sim m_q^2 \left(\frac{\alpha}{\pi}\right)^2 E^2~.
\label{nondiag_scales}
\end{equation}
Combining~\eqref{diag_scales} and~\eqref{nondiag_scales} then leads to a chirally-suppressed mixing angle for perturbative contributions
\begin{equation}
 \sin{2\theta}\sim \frac{\alpha}{\pi}\frac{m_q}{E}\ll 1~.
\label{mix_scales}
\end{equation}
A similar chiral suppression exists for the gluon condensate contributions
\begin{equation}
 \sin{2\theta}\sim \frac{m_q}{E}\frac{ \langle\alpha G^2\rangle}{E^4}\ll 1~.
\end{equation}
However, for the quark and mixed condensate contributions  we find
\begin{gather}
 \sin{2\theta}\sim \frac{\alpha}{\pi} \frac{\left\langle \bar q q\right \rangle}{E^3}
\label{qq_mix}
\\
\sin{2\theta}\sim \frac{\left\langle \bar q \sigma Gq\right\rangle}{E^5}=\frac{M_0^2}{E^2}\frac{\left\langle \bar q q\right \rangle}{E^3}~
.
\label{mix_mix}
\end{gather}
These estimates illustrate that the chiral-violating condensates avoid the chiral suppression of the mixing angle.  However, since $M_0\sim E$ the mixing angle generated by the quark condensate will be suppressed compared to the mixed condensate by a factor of  $\alpha/\pi$.   As noted earlier, the mixed condensate is zero at leading order in the non-diagonal pseudoscalar case \cite{paver} and hence there is a qualitative distinction between the scalar and pseudoscalar channels.  Although it is not as simple to estimate the order of magnitude of the mixing generated by the instanton contributions, one already sees from \eqref{mix_mix} that a substantive mixing angle is anticipated.

\begin{figure}[hbt]
\centering
\includegraphics[scale=0.5]{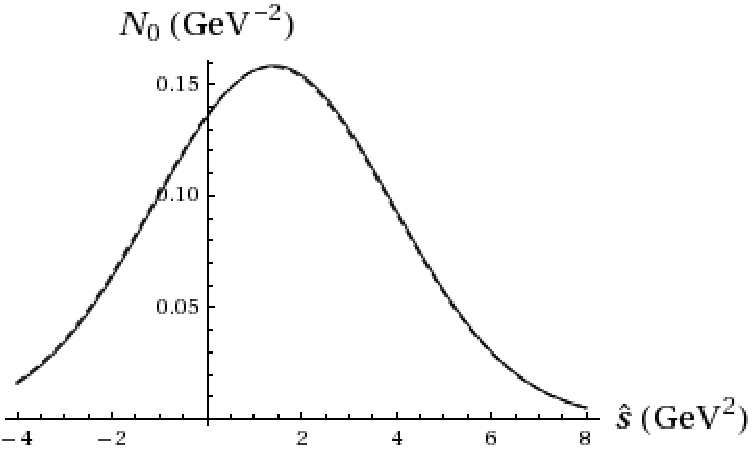}
\caption{
Comparison of the
 full contributions (solid curve) to the GSR $N_0^{(gq)}\left(\hat s, \tau,s_0\right)$ with its leading-order chiral contributions (dashed curve)
from the quark condensate, mixed condensate and instanton arising from $\chi_0^{(gq)}\left(\hat s,\tau,s_0\right)$.  The two curves overlap almost completely. Central values of the QCD input parameters have been employed along with $s_0=2.5\,{\rm GeV^2}$ and   $\tau=3\,{\rm GeV^4}$.
}
\label{nondiag_chiral_fig}
\end{figure}

The apparent contradiction between the strong mixing found in the GSRs for the diagonal correlators and the basic perturbative scales in the non-diagonal GSR is therefore resolved by a  detailed analysis of the non-diagonal case which demonstrates that the chiral-violating terms ({\it i.e.,} quark condensate, mixed condensate, and instanton) dominate the perturbative and gluon condensate corrections.  We first define the leading ${\cal O}\left(m_q\right)$   chiral terms in~\eqref{gq_GSR} as
\begin{equation}
 \begin{split}
\chi^{(gq)}_0\left(\hat s,\tau,s_0\right)&=-C_0m_q\langle\bar qq\rangle\frac{1}{\sqrt{4\pi\tau}}\int\limits_0^{s_0}dt\,
\exp{\left[-\frac{\left(t-\hat s\right)^2}{4\tau}\right]}
+D_0 m_q\left\langle\bar q\sigma G q\right\rangle \frac{1}{\sqrt{4\pi\tau}}\exp{\left(-\frac{\hat s^2}{4\tau}\right)}
\\
&+2\sqrt{3n_c}\pi m_q \rho^2 
\frac{1}{\sqrt{4\pi\tau}}
\int\limits_0^{s_0}dt\, \exp{\left[-\frac{\left(t-\hat s\right)^2}{4\tau}\right]}
t\sqrt{t}\left[
J_1\left(\rho\sqrt{t}\right)Y_2\left(\rho\sqrt{t}\right)+J_2\left(\rho\sqrt{t}\right)Y_1\left(\rho\sqrt{t}\right)
\right]~.
\end{split}
\label{N0_chiral}
\end{equation}
Figure \ref{nondiag_chiral_fig} demonstrates that these leading chiral terms are actually the dominant contribution to 
the non-diagonal GSR, avoiding the chiral suppression occurring in~\eqref{mix_scales}, and obviating the chiral suppression of the mixing angle that would occur for dominantly perturbative corrections.  We thus have the intriguing result that the underlying mixing mechanism is fundamentally non-perturbative, {\it i.e.,} perturbative analyses do not provide the essential phenomenological scales.

\begin{figure}[hbt]
\centering
\includegraphics[scale=0.5]{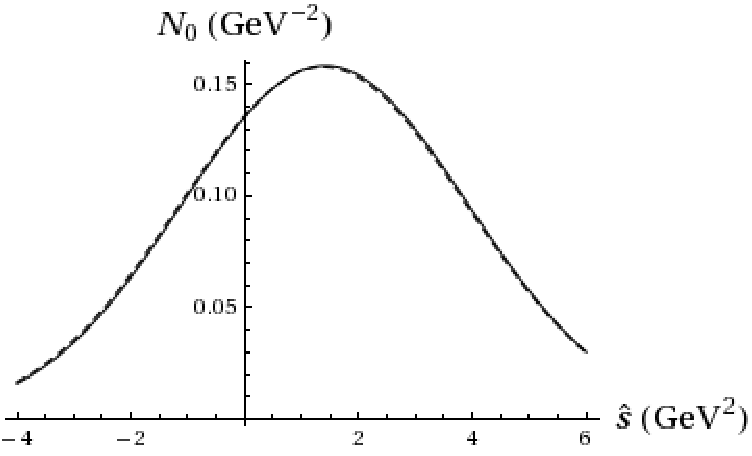}
\caption{
Comparison of the
  the NGSR $N_0^{(gq)}\left(\hat s, \tau,s_0\right)$  for the upper and lower ranges of the quark mass
specified in~\eqref{quark_mass}.  Central values of the other QCD input parameters have been employed along with $\tau=3\,{\rm GeV^4}$ and $s_0=2.5\,{\rm GeV^2}$.  The two curves overlap almost completely.
}
\label{N0_chiral_mq_fig}
\end{figure}

\begin{figure}[hbt]
\centering
\includegraphics[scale=0.5]{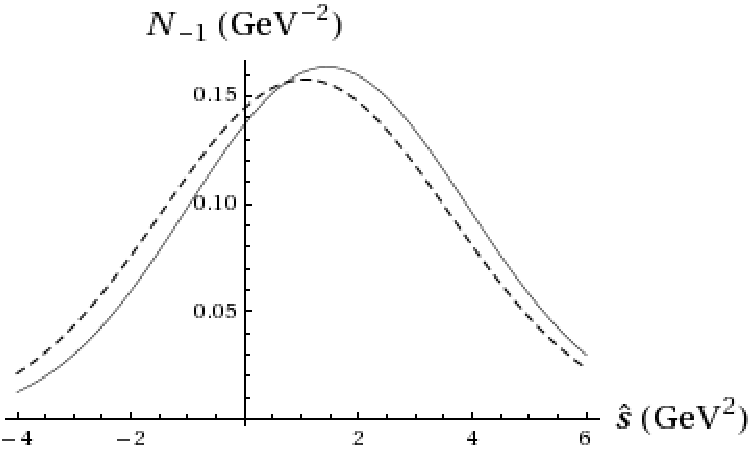}
\caption{
Comparison of the
  the NGSR $N_{-1}^{(gq)}\left(\hat s, \tau,s_0\right)$  for the upper and lower ranges of the quark mass
specified in~\eqref{quark_mass}.  Central values of the other QCD input parameters have been employed along with $\tau=3\,{\rm GeV^4}$ and $s_0=2.5\,{\rm GeV^2}$.  
 The solid and dotted curves respectively correspond to the lower  and upper bound on the quark mass.
}
\label{Nm1_chiral_mq_fig}
\end{figure}

Because PCAC fixes the combination $m_q\langle\bar qq\rangle$, the quark condensate term in~\eqref{N0_chiral} is effectively independent of $m_q$, so in principle $N_0^{(gq)}$ could be strongly dependent on $m_q$.  Fig.~\ref{N0_chiral_mq_fig} shows that this is not the case; the  NGSR $N^{(gq)}_0$ is relatively insensitive to the range~\eqref{quark_mass}
for $m_q$.  However, Fig.~\ref{Nm1_chiral_mq_fig} shows that this is not the case for the LET-sensitive NGSR
$N^{(gq)}_{-1}$ which exhibits stronger dependence on $m_q$.  Thus we focus our analysis on the $k=0$ NGSR
$N^{(gq)}_0$ as it is less affected by quark-mass uncertainties.  However, it is significant that 
Figs.~\ref{N0_chiral_mq_fig} and \ref{Nm1_chiral_mq_fig} do demonstrate qualitative agreement between $N_0$ and $N_{-1}$, particularly for $m_q$ at the upper bound of~\eqref{quark_mass}.     

\begin{figure}[hbt]
\centering
\includegraphics[scale=0.5]{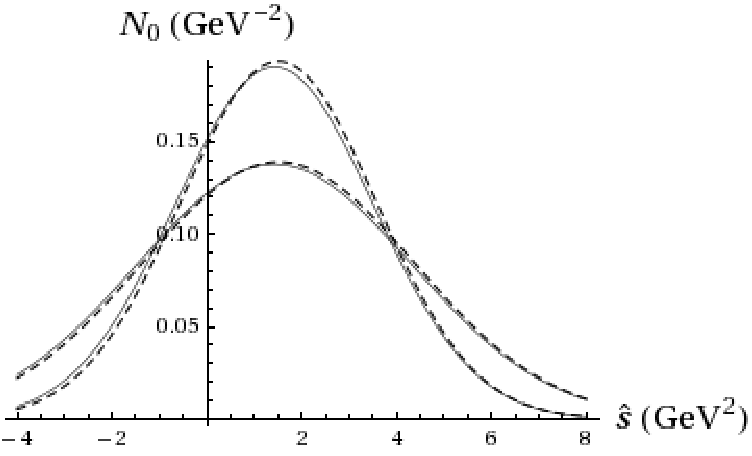}
\caption{
Comparison of the best fit of 
theoretical expression for the  normalized GSR $N_0^{(gq)}\left(\hat s, \tau,s_0\right)$ (solid curves) to the
double narrow resonance phenomenological model (dashed curves).  Resonance parameters resulting from the analyses of the diagonal NGSRs have been employed (the solution $r_1=0.45$ 
from~\eqref{r1_cases}, along with $m_1=0.98\,{\rm GeV}$ and $m_2=1.4\,{\rm GeV}$).  The optimized value of the 
  continuum for these (inputted) resonance parameters is $s_0=2.75\,{\rm GeV^2}$. 
 The upper set of curves are for $\tau=2\,{\rm GeV^4}$ and the bottom set of curves is for $\tau=4\,{\rm GeV^4}$. The phenomenological and QCD expressions  overlap  to a large extent.
}
\label{consistent_fig1}
\end{figure}

\begin{figure}[hbt]
\centering
\includegraphics[scale=0.5]{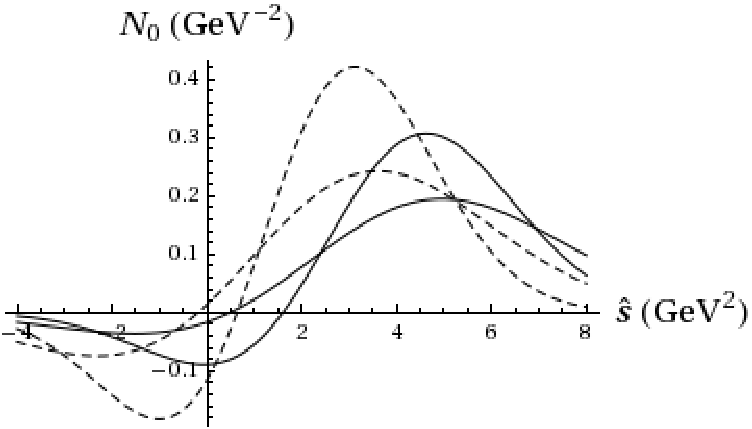}
\caption{
Comparison of the best fit of 
theoretical expression for the  normalized GSR $N_0^{(gq)}\left(\hat s, \tau,s_0\right)$ (solid curves) to the
double narrow resonance phenomenological model (dashed curves).  Resonance parameters resulting from the analyses of the diagonal NGSRs have been employed (the solution $r_1=-4.4$ 
from~\eqref{r1_cases}, along with $m_1=0.98\,{\rm GeV}$ and $m_2=1.4\,{\rm GeV}$).  The optimized value of the 
  continuum for these (inputted) resonance parameters is $s_0=4.9\,{\rm GeV^2}$. The upper set of curves are for $\tau=2\,{\rm GeV^4}$ and the bottom set of curves is for $\tau=4\,{\rm GeV^4}$. 
}
\label{consistent_fig2}
\end{figure}

Our detailed analysis of the non-diagonal GSR begins with an exploration of its consistency with the results of the diagonal cases.  In the double narrow resonance model, the non-diagonal NGSR has the form
\begin{gather}
 N^{(gq)}_0\left(\hat s,\tau,s_0\right)=\frac{1}{\sqrt{4\pi\tau}}
\left\{
  r^{(gq)}_1\exp\left[-\frac{(\hat{s}-m_1^2)^2}{4\tau}\right]
  +        r^{(gq)}_2\exp\left[-\frac{(\hat{s}-m_2^2)^2}{4\tau}\right]
\right\}~,
\label{nondiag_form}
\\
r^{(gq)}_1=\frac{f_{1g}f_{1q}}{f_{1g}f_{1q}+f_{2g}f_{2q}} ~,~ r^{(gq)}_2=\frac{f_{2g}f_{2q}}{f_{1g}f_{1q}+f_{2g}f_{2q}}~,~
r^{(gq)}_1+r^{(gq)}_2=1~.
\end{gather}
From the analysis of the diagonal cases \cite{harnett_glue,harnett_quark,harnett_proc}, we have found
(see Table~\ref{doubres_tab}) $m_1\approx 1\,{\rm GeV}$, $m_2\approx 1.4\,{\rm GeV}$, and
\begin{gather}
r^{(gg)}_1=0.28=\frac{{f_{1g}}^2}{{f_{1g}}^2+{f_{2g}}^2}~,~r^{(gg)}_2=1-0.28=\frac{{f_{2g}}^2}{{f_{1g}}^2+{f_{2g}}^2}
\label{gg_couplings}
\\
r^{(qq)}_1=0.63=\frac{{f_{1q}}^2}{{f_{1q}}^2+{f_{2q}}^2}~,~r^{(qq)}_2=1-0.63=\frac{{f_{2q}}^2}{{f_{1q}}^2+{f_{2q}}^2}~.
\label{qq_couplings}
\end{gather}
Thus the parametrization of the mixed gluonic-$\bar q q$ system has four couplings of the states to the various currents as in Ref.~\cite{ailin}.  The  four equations~\eqref{gg_couplings} and~\eqref{qq_couplings} representing the diagonal results determine the four couplings up to an overall sign, leading to two possible solutions for the non-diagonal case:
\begin{equation}
 r^{(gq)}_1=\begin{cases}+0.45\\-4.4\end{cases}~.
\label{r1_cases}
\end{equation}
Apart from the ambiguity arising from the sign of the couplings, all the phenomenological parameters in the non-diagonal NGSR~\eqref{nondiag_form} are determined except for the continuum $s_0$ which can be determined by performing a least-squares fit of the $\hat s$, $\tau$ dependence of \eqref{nondiag_form} in the region
 $-4\,{\rm GeV^2}<\hat s<8\,{\rm GeV}^2$ and $2\,{\rm GeV^4}<\tau<4\,{\rm GeV}^4$.  The best fit for the two cases in~\eqref{r1_cases} are shown in Figs.~\ref{consistent_fig1} and~\ref{consistent_fig2}. From these Figures we see that  the positive case in~\eqref{r1_cases} is demonstrably most consistent with the QCD expression.  

At this point we  reach  an important conclusion: the  non-diagonal GSR is consistent with the 
results of the diagonal gluonic GSR analyses \cite{harnett_glue,harnett_quark,harnett_proc}, providing strong evidence for  a consistent scenario of mixed gluonic-$\bar q q$ states with masses of $m_1\approx 1\,{\rm GeV}$ and $m_2\approx 1.4\,{\rm GeV}$ that couple to mixtures of gluonium and $q\bar q$ currents,
with the heavier state having a slightly larger gluonic coupling.   
A significant feature of our analysis is the nearly-identical masses that have resulted independently from 
the diagonal gluonic and diagonal $\bar q q$ cases.  Although we have demonstrated that the non-diagonal case is consistent with these results, an independent analysis of the non-diagonal case is necessary for further confirmation of  this gluonic-$\bar q q$ mixing scenario.

\begin{figure}[hbt]
\centering
\includegraphics[scale=0.5]{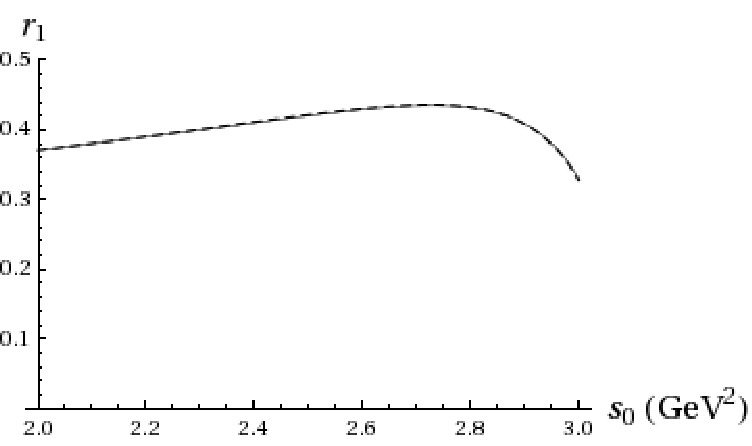}
\caption{
The resonance parameter $r^{(gq)}_1$ extracted from the moments \eqref{r_moms} of $N_0^{(gq)}$ as a function of $s_0$ for $\tau=2\,{\rm GeV^4}$ (solid curve) and  $\tau=4\,{\rm GeV^4}$ (dashed curve).  The dashed and sold curves overlap almost completely.
}
\label{r_fig}
\end{figure}

\begin{figure}[hbt]
\centering
\includegraphics[scale=0.5]{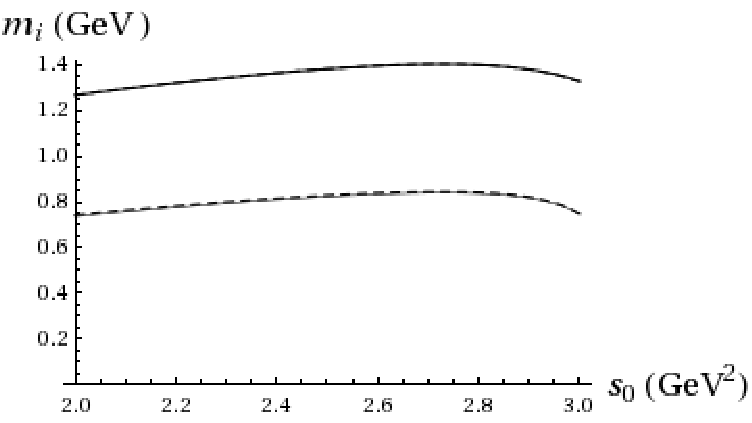}
\caption{
The resonance parameters $m_1$ and $m_2$  extracted from the moments \eqref{z_moms} and \eqref{y_moms} 
of $N_0^{(gq)}$ as a function of $s_0$ for $\tau=2\,{\rm GeV^4}$ (solid curve) and  $\tau=4\,{\rm GeV^4}$ (dashed curve).  The dashed and sold curves overlap almost completely in both cases. The upper pair of curves represent $m_2$ and the lower set represent $m_1$.
}
\label{mass_fig}
\end{figure}

The moments \eqref{z_moms}--\eqref{r_moms} will be used to determine the QCD predictions of the resonance parameters arising from the non-diagonal NGSR $N_0^{(gq)}$.  In Figs.~\ref{r_fig} and \ref{mass_fig} we show the $s_0$, $\tau$ dependence of these moments. As required in a reliable sum-rule analysis,  the resulting resonance parameters show almost no $\tau$ dependence.   The predicted value of the resonance parameters is then obtained from the point where they are stable against variations in $s_0$.  This point of stability is approximately $s_0=2.75\,{\rm GeV^2}$ in all cases and is comparable in scale to that resulting from the diagonal analysis (see Table~\ref{doubres_tab}), confirming the reliability of the procedure.  The results of the analysis of the non-diagonal 
NGSR are given in Table~\ref{doubres_tab}. Although the theoretical uncertainties for the non-diagonal sum-rule have not yet been discussed, it is evident that the independent predictions of the masses from all possible sum-rules show exceptional agreement, particularly for the heavier state.  Furthermore, the prediction $r_1^{(gq)}=0.44$ from the non-diagonal sum-rule is in excellent agreement with the positive solution of Eq.~\eqref{r1_cases}  emerging from the diagonal sum-rules.

\begin{figure}[hbt]
\centering
\includegraphics[scale=0.5]{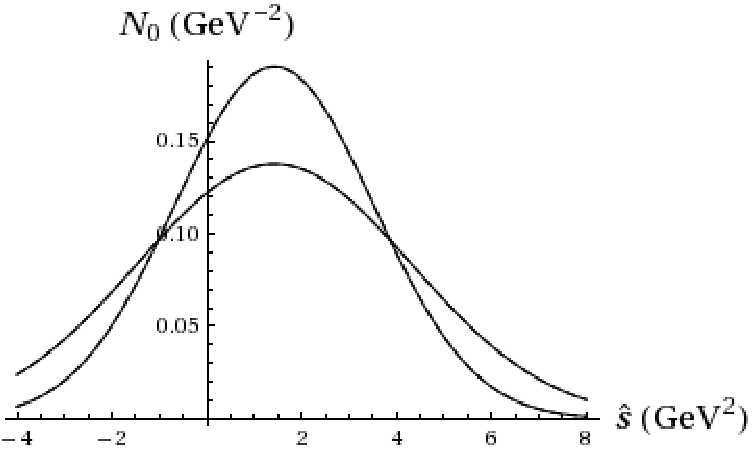}
\caption{
Comparison of the
theoretical expression (solid curves) for the non-diagonal NGSR $N_0^{(gq)}\left(\hat s, \tau,s_0\right)$ with the
double narrow resonance phenomenological model (dashed curves) for the predicted values of the resonance parameters and continuum $s_0$ of Table \protect\ref{doubres_tab}.  
 The upper set of curves are for $\tau=2\,{\rm GeV^4}$ and the bottom set of curves is for $\tau=4\,{\rm GeV^4}$.
The dashed and solid curves overlap almost completely in both cases. 
}
\label{nondiag_fit_fig}
\end{figure}

\begin{figure}[hbt]
\centering
\includegraphics[scale=0.5]{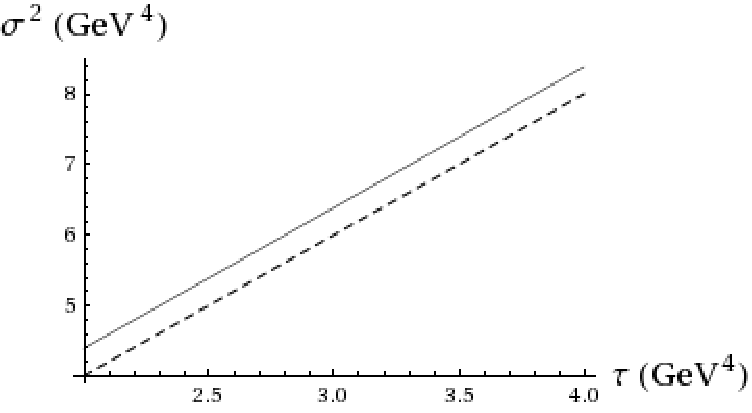}
\caption{
Comparison of the theoretical value of the moment $\sigma^2\left(\tau,s_0\right)$ (solid curve) with $2\tau$ (dashed curve).  The optimized value $s_0=2.75\,{\rm GeV^2}$ has been used along  with central values of QCD input parameters.
}
\label{sig_fig}
\end{figure}

For the non-diagonal NGSR, the double narrow resonance model results in excellent agreement with the QCD expression as illustrated in Figure \ref{nondiag_fit_fig};
as in the diagonal case (see Fig.~\ref{twores_fig}) there is no indication of discrepancies  that would require a more elaborate phenomenological model ({\it e.g.,} additional states, resonance widths).  As a further diagnostic, Fig.~\ref{sig_fig} illustrates that the optimized value $s_0=2.75\,{\rm GeV^2}$ leads to excellent agreement between the $\tau$ dependence of the QCD expression $\sigma^2$ and its expected dependence \eqref{sig_rel} from the two resonance model:  a straight line with slope $2$ and a positive intercept.

Because the non-diagonal sum-rule is dominated by the chiral-violating contributions, the Table~\ref{doubres_tab} predictions of the resonance parameters  only depend on the quark mass, mixed condensate, instanton size, and instanton density.   The quantity $r_1^{(gq)}$ is most dependent on $\hat m_q$ and $\rho$, whereas $m_1$ and $m_2$ are most dependent on $\rho$.   By comparison, the resonance parameters are relatively unaffected by uncertainties in the mixed condensate and $n_c$.   In aggregate, the resulting uncertainties in the mass parameters are comparable to those found in the diagonal case: approximately $0.2\,{\rm GeV}$ with a correlated effect that leads to a relatively stable mass splitting $0.5\,{\rm GeV}<m_2-m_1< 0.6\,{\rm GeV}$.  However, in comparison to the diagonal analyses, the couplings have greater sensitivity to the input parameters in the non-diagonal case, with an uncertainty in $r_1^{(gq)}$  of approximately $0.1$.  Across the entire parameter space considered, $s_0$ continues to stabilize at the same value for all the resonance parameters, demonstrating that our methodology is robust.

Taking into account the uncertainties in the values of the resonance parameters for the diagonal and non-diagonal NGSRs associated with Table~\ref{doubres_tab}, we see that the non-diagonal case leads to predictions that are consistent with those of the diagonal analyses, so that all possible GSRs of gluonic and $\bar q q$ currents independently confirm 
the existence of two states with approximate masses of $1\,{\rm GeV}$   and $1.4\,{\rm GeV}$ 
that couple to 
mixtures of gluonium and $q\bar q$ currents.  In particular, there is excellent agreement between the central value $r_1^{(gq)}=0.44$ obtained from analysis of the non-diagonal NGSR and the positive solution $r_1^{(gq)}=0.45$ in Eq.~\eqref{r1_cases} resulting from the diagonal NGSRs.
 
Another inherent source of uncertainty in our analysis is the narrow resonance approximation.  One can qualitatively model the effect of resonance widths and kinematic distortions by studying their effect on the moments \eqref{first_moment}--\eqref{dist_asymm}.  For example,  a Gaussian resonance (which introduces an effective Breit-Wigner width $\Gamma_{BW}$) with a $t^2$ kinematic distortion \cite{liu,dominguez} can be modelled by decreasing the QCD values of the moments systematically as a function of the ratio $\xi=\sqrt{2\log{2}}\Gamma_{BW}/m$ \cite{harnett_glue} and then exploring the effect on the resonance parameters as a series in $\xi$.  The leading-order deviations of $m_1$  and $m_2$ from their narrow-width values are proportional to $\xi^2$ with a negative coefficient of ${\cal O}(1)$ in GeV units.  From this we can conclude that the narrow-width approximation tends to overestimate the masses if the underlying resonances are broad and kinematically-skewed  (see {\it e.g.,} \cite{vic} for similar conclusions in other contexts).  
 In this situation, the  mass predictions in Table~\ref{doubres_tab} can be conservatively interpreted as an upper bound on the masses in a more complicated models \cite{liu,dominguez}.

\begin{figure}[hbt]
\centering
\includegraphics[scale=0.5]{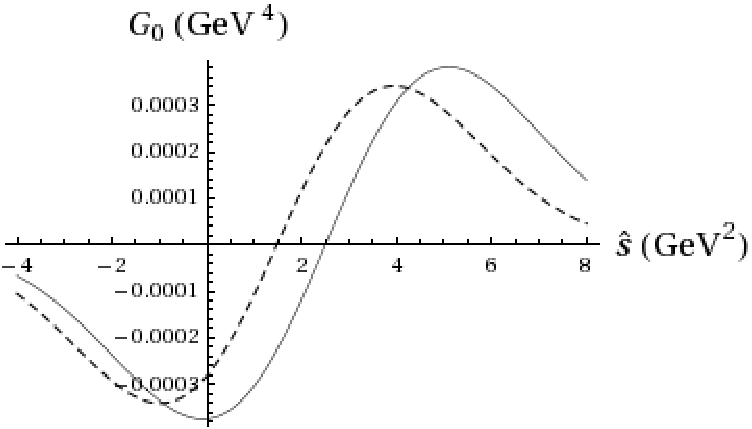}
\caption{
Comparison of the best fit of 
theoretical expression for the  GSR $G_0^{(gq)}\left(\hat s, \tau,s_0\right)$ to the
double narrow resonance phenomenological model for a single mixing-angle.  Resonance masses $m_1=1\,{\rm GeV}$ and $m_2=1.4\,{\rm GeV}$ resulting from the analyses of the diagonal NGSRs have been employed. The 
  continuum leading to $M_{0,0}^{(gq)}=0$ for the central values of QCD input parameters 
is $s_0=4.18\,{\rm GeV^2}$. 
 The solid  curve represents the QCD expression $G_0^{(gq)}\left(\hat s, \tau,s_0\right)$
and the dashed curve represents the phenomenological model; $\tau=3.0\,{\rm GeV^4}$ has been used in both cases. 
}
\label{equal_strength_fig}
\end{figure}

We now examine the pattern of mixing of the couplings for these states as contained in the quantities $r_i$ for the various cases.  
If the couplings obey a single-angle mixing pattern, then one would expect $r_2^{(gg)}=\cos^2\theta=r_1^{(qq)}$.  Although such a scenario could be possible given the 10\% uncertainty in these quantities in Table~\ref{doubres_tab},  the non-diagonal case provides a more sensitive test because single-angle mixing leads to 
\begin{gather}
 G^{(gq)}_0\left(\hat s,\tau,s_0\right)\sim\frac{\sin{2\theta}}{\sqrt{4\pi\tau}}
\left[
  \exp\left[-\frac{(\hat{s}-m_1^2)^2}{4\tau}\right]
          -\exp\left[-\frac{(\hat{s}-m_2^2)^2}{4\tau}\right]
\right]~.
\label{single_angle_form}
\end{gather}
In this situation, the integral of the right-hand side of~\eqref{single_angle_form} is zero which then requires $M_{0,0}^{(gq)}\left(\tau, s_0\right)=0$ in~\eqref{N_gq}. A value of $s_0=4.18\,{\rm GeV^2}$ can be found to satisfy this constraint over the considered range $2\,{\rm GeV^4}<\tau<4\,{\rm GeV^4}$.  However, as shown in Fig.~\ref{equal_strength_fig} the $\hat s$, $\tau$ dependence of the QCD expression $G^{(gq)}_0\left(\hat s,\tau,s_0\right)$ is not consistent with  mass scales $m_1\approx  1\,{\rm GeV}$ and $m_2\approx 1.4\,{\rm GeV}$.  We thus conclude that the pattern of mixing for the couplings is not consistently described by  a single mixing angle, and hence the situation must be similar to the two-angle scenario that has been found for the couplings (decay constants) of the $\eta$-$\eta'$ system in the singlet-octet basis \cite{feldmann}.  Implicitly this is the same result found in \cite{lattice2,ailin}, where four independent couplings are found necessary in the study of the mixed gluonic-$q\bar q$ system rather than the three-parameter system of two couplings and one mixing angle.  Following Ref.~\cite{lattice2}, we then define an effective mixing angle $\phi$ 
\begin{equation}
\tan^2\phi=\left\vert
\frac{\langle 0\vert J_g \vert 1\rangle\, \langle 0\vert J_q\vert 2\rangle}{\langle 0\vert J_g\vert 2\rangle\, \langle 0\vert J_q\vert 1\rangle} \right\vert~,
\end{equation}
where $\vert 1\rangle$ and $\vert 2\rangle$ respectively correspond to the states with mass $m_1$ and $m_2$.
Taking into account the uncertainties in the Table~\ref{doubres_tab} values  leads to  $\phi=54^\circ\pm 4^\circ$.  Thus we find that the effective mixing angle corresponds to nearly maximal mixing ($\phi=45^\circ$) where each of the two  states are equally coupled to the $q\bar q$ and gluonic currents, a result in excellent agreement with the conclusions of \cite{narison_review,narison}.  The deviation of our effective mixing angle from the maximal angle indicates that the
heavier ($1.4\,{\rm GeV}$) state $\vert 2\rangle$ is somewhat more gluonic in comparison to the lighter
($1.0\,{\rm GeV}$) state $\vert 1\rangle$.

\section{Discussion and Conclusions}
Gaussian QCD sum-rules are able to probe hadronic spectral functions over a broad range of energy, and are thus ideally suited to exploring the possibility of states that couple to mixtures of gluonium and $q\bar q$  currents exist amongst the light scalar mesons.  We have   studied the NGSRs for all possible combinations of scalar gluonic and scalar $I=0$ (non-strange) $q\bar q$ currents (diagonal gluonic, diagonal $q\bar q$, and non-diagonal $q\bar q$-gluonic) and  find that all three cases independently predict the existence of 
two states with masses of approximately 
$1\,{\rm GeV}$  and  
$1.4\,{\rm GeV}$.  This is precisely what one would expect from hadronic states that couple to mixtures of gluonium and $q\bar q$ currents.    Given the uncertainties in our mass predictions, it is not clear whether our lighter state should be interpreted as the $f_0(980)$  or $\sigma$ (at the heavier end of its range \cite{pdg}) and it is also not clear whether the heavier state should be interpreted as the $f_0(1370)$ or the $f_0(1500)$.  However, because the approximate $0.5\,{\rm GeV}$ mass splitting between the states is relatively stable under QCD uncertainties, our results do suggest identifying either the lighter pair of states [$\sigma$ and $f_0(1370)$] or the
heavier pair [$f_0(980)$ and $f_0(1500)$]  as states coupling to mixtures of gluonium-$q\bar q$ currents.

The non-diagonal sum-rule provides important insights into the mixing of the $q\bar q$ and gluonic aspects of  these two states.  Because of chiral suppression factors associated with the light (non-strange) quarks, perturbative effects are unable to generate any significant amount of mixing.  However, the chiral-violating contributions of the quark condensate, mixed condensate,  and instantons do not suffer from this chiral suppression and provide the dominant contribution to the non-diagonal correlation function, implying that mixing of gluonic and $q\bar q$ degrees of freedom has a non-perturbative origin.   Qualitatively, this conclusion is similar to that obtained for glueball decays \cite{hongying} and to that of Ref.~\cite{kochelev}  which  demonstrated that instantons can lead to a significant mixing between glueballs and (heavy quark) mesons in the pseudoscalar channel.

The state couplings that result from the analysis of the various GSRs  provide an additional means to examine the self-consistency of the scenario of
two states with masses of approximately
 $1\,{\rm GeV}$ and 
$1.4\,{\rm GeV}$   that couple to mixtures of $q\bar q$ and gluonic currents.  In particular, the relative couplings between the states 
in the non-diagonal case is constrained by the relative couplings in the diagonal cases.  The independent prediction of these couplings from the non-diagonal NGSR is found to satisfy this constraint extremely well, providing strong evidence for the validity of the mixing scenario.

The state couplings also provide a means to study the pattern of mixing associated with the couplings to gluonic and $q\bar q$ currents.  The resulting pattern is similar to the two-angle  mixing that occurs for the couplings (decay constants) for the $\eta$-$\eta'$ system  in the singlet-octet basis \cite{feldmann}, and result in an effective mixing angle of $\phi\approx 54^\circ$ in excellent agreement with the sum-rule analyses of Refs.~\cite{narison_review,narison}.  Because this mixing angle is in the region near maximal mixing ($\phi=45^\circ$), there is only a slight preference for the heavier $1.4\,{\rm GeV}$ state to couple to gluonic currents and a concomitantly slight preference for the lighter $1.0\,{\rm GeV}$ state to couple to $q\bar q$ currents.    Indeed, the existence of such strong mixing implies that qualitative features that would distinguish pure gluonic and $q\bar q$ states 
would be obscured for strongly-mixed states and the experimental signal of gluonium would thus be  elusive.

In summary, our results provide strong QCD evidence to support  the scenario where the mixing of $q\bar q$ and gluonium  is manifested in the scalar hadronic spectrum as a lighter state on the order of $1\,{\rm GeV}$ and
 a heavier state on the order of $1.5\, {\rm GeV}$ 
 \cite{amir,minkowski,lattice2,narison_review,narison,ailin,kisslinger,harnett_proc}. 
In particular,  our conclusion that  there exists a strong  mixing between gluonium and $q\bar q$  states is similar to the results from a variety of approaches~\cite{minkowski,lattice2,narison_review,narison} and our result for the heavier state's  preference for gluonic channels  provides QCD support for the findings of a large gluonic component of the $f_0(1500)$ \cite{close,f0_1500_interp}.

\smallskip
\noindent
{\bf Acknowledgements:}  The authors  are grateful for financial support
from the Natural Sciences and Engineering Research Council of Canada (NSERC).  Many thanks to Hong-Ying Jin for helpful discussions. 
TGS dedicates this work to the memory of Victor Elias.

\end{document}